\documentclass[11pt,hyper]{JHEP3}
\usepackage{graphicx,amssymb,amsmath,amsfonts,cite}
\usepackage[english]{babel}
\usepackage{amssymb}
\usepackage{amsfonts}
\usepackage{amsmath}
\usepackage{graphicx}
\usepackage{epsfig}
\usepackage{cite}
\newcommand{\cC}{{\cal C}}  
\newcommand{\cE}{{\cal E}}

  \newcommand{\cL}{{\cal L}}

\newcommand{\tA}{{\tilde{A}}}
\newcommand{\tF}{\tilde{F}}
\newcommand{\tE}{{\tilde{E}}}
\newcommand{\tC}{{\tilde{C}}}

\newcommand{\be}{\begin{equation}} \newcommand{\ee}{\end{equation}}
\newcommand{\bea}{\begin{eqnarray}} \newcommand{\eea}{\end{eqnarray}}
\newcommand{\beann}{\begin{eqnarray*}}  \newcommand{\eeann}{\end{eqnarray*}}
\newcommand{\bfig}{\begin{figure}} \newcommand{\efig}{\end{figure}}
\newcommand{\ba}{\begin{array}} \newcommand{\ea}{\end{array}}
\newcommand{\bcen}{\begin{center}} \newcommand{\ecen}{\end{center}}
\newcommand{\btab}{\begin{tabular}} \newcommand{\etab}{\end{tabular}}

\newcommand{\bra}[1]{\langle #1|}
\newcommand{\ket}[1]{|#1\rangle}
\newcommand{\vev}[1]{\left\langle{#1}\right\rangle}

\newtheorem{Proposition}{Proposition}[section]

\newtheorem{Theorem}{Theorem}[section]
\newtheorem{Lemma}{Lemma}[section]

\newcommand{\bp}{\begin{Proposition}}   \newcommand{\ep}{\end{Proposition}}
\newcommand{\bt}{\begin{Theorem}}   \newcommand{\et}{\end{Theorem}}
\newcommand{\bl}{\begin{Lemma}}     \newcommand{\el}{\end{Lemma}}
\newcommand{\bc}{\begin{Corolary}} \newcommand{\ec}{\end{Corolary}}
\def\bra{\left\langle}
\def\ket{\right\rangle}
\def\vev#1{{\bra #1 \ket}}
\def\pa{\partial}
\def\CR{\nonumber\\}

\def\d{\partial}

\title{Stability conditions for spatially modulated phases}

\author{Sophia K. Domokos${}^1$, Carlos Hoyos${}^2$, Jacob Sonnenschein${}^2$,
\\
${}^1$\textit{Weizmann Institute of Science, Rehovot 76100, Israel}\\
\\
${}^2$ \textit{
  Raymond and Beverly Sackler Faculty of Exact Sciences \\
School of Physics and Astronomy \\
Tel-Aviv University, Ramat-Aviv 69978, Israel.}\\
E-mail: \tt{ sophia.domokos@weizmann.ac.il, choyos@post.tau.ac.il, cobi@post.tau.ac.il}
}

\abstract{We introduce a novel set of stability conditions for vacua
with broken Lorentz symmetry. The first class of conditions require that the energy be minimized under small geometric deformations, which translates into requiring the positivity of a ``stiffness'' four-tensor. The second class of conditions requires that stress forces be restoring under small deformations.   We then apply these conditions to examples of recently-discovered spatially modulated (or ``striped'') phases  in holographic models of superconductors and high-density
QCD. For backreacted solutions we find that the pressure condition is equivalent to thermodynamic stability. For probe solutions, however, these conditions are in conflict with the minimization of the free energy. This suggests that either the solutions are unstable or the definition of the free energy in the probe approximation  must be revised for these solutions.}

\preprint{TAUP-2970/13, WIS/08/13-JUL-DPPA}

\begin{document}

\section{Introduction}

In systems at finite density, the energetically-preferred vacuum may break translational
and rotational invariance.
For instance, in large-$N_c$ QCD a chiral density wave is the
ground state
at low temperatures \cite{Deryagin:1992rw,Bringoltz:2009ym}\footnote{Although for $N=3$ in four dimensions the ground state is expected to be a color superconductor \cite{Shuster:1999tn}.}, and in the presence of a magnetic field a chiral magnetic wave \cite{Kharzeev:2010gd} or a chiral magnetic spiral \cite{Basar:2010zd} can form. Inhomogeneous phases are also important in condensed matter systems, where they may prove relevant for the description of high-$T_c$ superconductors \cite{larkin:1964zz,Fulde,Vojta}.

Instabilities of the spatially homogeneous vacuum abound in
holographic duals of strongly coupled systems. First identified in
\cite{Domokos:2007kt} for zero-temperature, finite-density QCD,
the authors in \cite{Nakamura:2009tf} provided a more general treatment, noting
that Chern-Simons-like terms may generically induce spatially
modulated instabilities for sufficiently large values of the
density or of the Chern-Simons coupling. These
instabilities have since been found in a variety of models, from the
probe-limit $AdS_5$ Reissner-Nordstrom black hole
\cite{Nakamura:2009tf} and the Sakai-Sugimoto model at zero
temperature and large axial chemical potential \cite{Bayona:2011ab}, to
duals of Fermi liquids in the probe limit
\cite{Bergman:2011rf,Iizuka:2013ag} and beyond \cite{Iizuka:2013ag},
as well as in broad classes of gravity duals for superconducting
materials (\cite{Donos:2011ff, Donos:2012gg, Donos:2011qt,Donos:2011pn,
Donos:2012yu}). \cite{Donos:2013gda} showed, furthermore, that duals
of charge density waves can arise even
without the parity- and time-translation-invariance-violating Chern-Simons
interaction.
Meanwhile, factors which can stabilize the homogeneous vacuum even
in the presence of the Chern-Simons coupling include $R^2$
gravitational corrections \cite{Takeuchi:2011uk}, magnetic fields
\cite{Jokela:2012vn,BallonBayona:2012wx}, and coarse-grain
holographic models of quarkonium \cite{deBoer:2012ij}.\footnote{It
is important to note that the spatially-modulated
instability can occur for solutions to the full $D=11$ supergravity
equations of motion of string- or M-theory (see e.g.
\cite{Donos:2011bh}): though it can be disrupted by varying additional parameters, it does not seem to be an artifact of
bottom-up or probe-limit models.}

The actual striped vacuum state has been explicitly constructed in
holographic QCD at high temperature and density
\cite{Ooguri:2010kt,Ooguri:2010xs}, at zero temperature and large
axial chemical potential \cite{Bayona:2011ab}, in supergravity duals
of superconductors
\cite{Donos:2012gg,Donos:2012wi,Donos:2013wia,Rozali:2013ama,Withers:2013kva,
Iizuka:2012pn,Rozali:2012es}. Though examples of striped phases are
quite common,\footnote{A partial classification of possible phases
was given in \cite{Iizuka:2012pn}.} these constitute just one of
many types of inhomogeneus vacua. Examples of solitons and more complicated
inhomogeneous phases include
\cite{Albash:2008eh,Albash:2009ix,Albash:2009iq,Montull:2009fe,Keranen:2009vi,Maeda:2009vf,Bu:2012mq,Kaplunovsky:2012gb,Kaplunovsky:2013iza}.

Though models with explicitly broken translation invariance are also
of great interest (e.g. for modeling lattice effects in condensed
matter systems
\cite{Hartnoll:2012rj,Horowitz:2012ky,Horowitz:2012gs,Donos:2012js,Liu:2012tr,
Vegh:2013sk}), in this note we focus exclusively on systems with
spontaneously broken translation invariance.

We can gain some intuition about inhomogeneous solutions from solid
state physics. A crystalline solid is an example of
an inhomogeneous state: the atoms are arranged in a periodic
structure that at large distances looks like a continuous medium.
The translational symmetries are spontaneously
broken, with the phonons acting as Goldstone bosons.
Deformations of the solid displace atoms from their equilibrium
positions, costing a finite amount of energy. When slightly deformed, such a
solid may experience internal stress forces that tend to bring
it back to its original shape: such deformation are {\em
elastic}. Large deformations which change the
shape of the solid permanently are termed {\em plastic}.

With this framework in mind, we study the stability of inhomogeneous solutions under small deformations. Our strategy differs from the usual linear stability analysis,
in which one studies time-dependent harmonic perturbations close to the static solution in configuration space. If the solution remains close to the original solution for all times (i.e. it represents oscillations around a minimum of the energy), it is declared stable. This strategy often involves solving highly non-trivial second order partial differential equations (PDEs). We, however, deform the static solution to another static configuration, still close to the original solution. In general the deformed configuration is not a solution of the equations of motion, but one can interpret it as the result of applying small external forces. If the energy of the deformed configuration is lower than the original solution, it is clear that when the forces are turned off, either suddenly or
gradually, the evolution of the system will not take it back to the original solution, but rather to some
different vacuum.

We develop two different types of stability conditions. The first uses the changes
in the system's energy due to small geometrical deformations, while the second uses changes in
the momentum. In the first case,  we expand
the energy to second order in the deformation and determine whether
 the inhomogeneous solution is truly
a minimum of the energy, or whether there are unstable directions.
The second set of conditions evaluates the stress forces on deformed solutions
using the energy-momentum tensor. The latter approach was used for
instance by Gibbons to study multiple BIon solutions
\cite{Gibbons:1997xz}. We will apply these
checks of stability to the spatially modulated solutions found by
Ooguri and Park in a Maxwell-Chern-Simons model \cite{Ooguri:2010kt}
and in the Sakai-Sugimoto model \cite{Ooguri:2010xs}. We do not
observe any sign of instability from the condition of energy
minimization, but we find the stress force condition to be in
tension with the condition of thermodynamic stability obtained from
minimizing the free energy. However, we find that a similar analysis
of backreacted solutions (using recent results of Donos and
Gauntlett \cite{Donos:2013cka}), reveals that in such cases the
two conditions are equivalent. This suggests either that  the definition of the free
energy in the probe limit is inconsistent, or that the backreaction is crucial to stabilizing the solutions.

Though we apply our minimal energy condition specifically to solutions periodic along
a spatial direction, most of our results apply generally to  static
solutions. The minimum energy condition, in particular, applies also
to soliton (finite energy) solutions, an extension of our previous work
\cite{Domokos:2013xqa}, where we derived conditions on the existence
of solitons from first-order deformations.

The paper is organized as follows: In section \S~\ref{sec:elastic}
we derive the minimum energy condition for generic theories with
scalar or gauge fields, and compare it to stability conditions in
elasticity. In section \S~\ref{sec:stress} we find a condition on
the variation of the pressure with respect to the period of the
spatially modulated solutions using a stress force analysis. In
section \S~\ref{sec:spatiallymodulated} we apply the stability
conditions from the previous sections on striped solutions. In
section \S~\ref{sec:end} we conclude and suggest several directions
for future work. We have gathered some calculational details and
additional material in two appendices.

\section{Minimum energy condition}\label{sec:elastic}
Consider a static solution to the equations of motion in a given field theory. Now perform a geometrical transformation -- such as a shear or dilation --
on the solution.
A solution is energetically stable if its energy increases (or stays the same) under any such deformation. This requirement leads
to a series of constraints on static stable solutions.

In what follows we will consider for simplicity only scalar fields
or Abelian vector fields. The extension to more generic cases is
straightforward.

\subsection{Deformed configurations to first order}

In this section we review some of the results of
\cite{Domokos:2013xqa} in order to introduce our methodology and
establish some notation. Consider the simple case of a theory with
one or more scalars $\phi^a$, that possesses a static solution,
$\phi^a_0(x)$ with finite energy density. How do small deformations
affect the energy of this solution?

Generically, the energy of the solution is a function of the fields
and its derivatives,
\begin{equation}
E[\phi^a_0]=\int d^dx\,\cE(\phi^a_0,\partial_i\phi^a_0),
\end{equation}
where $\cE$ is the energy density. Let's say we deform the
configuration by a geometric deformation in space, $\Lambda x$. The
deformed solution takes the form
\begin{equation}
\phi^a_\Lambda(x)=\phi^a_0(\Lambda x)~.
\end{equation}
For small deformations we can do an expansion around the undeformed solution:
\begin{equation}\label{eq:defxi}
(\Lambda x)^i\simeq x^i+\xi^i(x)~.
\end{equation}
We will use Latin indices for spatial components
$i,j,\cdots=1,\dots,d$ and Greek indices for spacetime components
$\mu,\nu,\cdots=0,1,\dots,d$. The energy of the deformed
configuration is, to leading order,
\begin{align}
\notag E[\phi^a_\Lambda] &=\int d^dx\,\cE\left(\phi^a_\Lambda(x),\partial_i\phi^a_\Lambda(x)\right)\\
\notag &=\int d^d \tilde{x} \left|\left|\frac{\delta x^i}{\delta {\tilde x^j}}\right|\right|\cE\left(\phi^a_0(\tilde{x}),\frac{\partial {\tilde x^j}}{\partial x^i}\tilde{\partial}_j \phi^a_0(\tilde{x})\right)\\
\notag &\simeq \int d^d \tilde{x}\cE\left(\phi_0^a(\tilde{x}),\tilde{\partial}_i \phi^a_0(\tilde{x})\right)+\int d^d \tilde{x}\, \tilde{\partial}_i\xi^j\left[\delta^i_{\ j}\cE-\frac{\delta \cE}{\delta \tilde{\partial}_i \phi^a}\tilde{\partial}_j\phi^a_0\right]\\
&=E[\phi^a_0]-\int d^d \tilde{x}\, \tilde{\partial}_i\xi^j \Pi^i_{\
j}(\phi^a_0)+\dots.
\end{align}
In the second line we have made the change of variables
${\tilde x^i}=(\Lambda x)^i$ and in the next lines we have expanded for
$x^i\simeq {\tilde x^i}-\xi^i$. Here $\Pi^i_{\ j}$ is a stress tensor for
static configurations based on the energy density $\cE$:
\begin{align}
\Pi^i_{\ j}=\frac{\delta \cE}{\delta \d_i\phi^a}\d_j\phi^a-\delta^i_j\cE~.
\end{align}
The difference in the energy of the deformed solution compared to
the original one is given by the stress tensor ($\Pi^i_{\ j}$) evaluated
at the soliton solution:
\begin{equation}
E[\phi^a_\Lambda]-E[\phi^a_0]=\int d^d x \ \delta\cE=-\int d^d x\,
\partial_i\xi^j \Pi^i_{\ j}.
\end{equation}
For a static solution to the equations of motion this term must be a
total derivative.\footnote{For scalars this is obvious, as
$\cE=-\cL$. For theories with gauge fields the situation is a bit
more complicated, as the energy density and stress tensor include
improvement terms; however, the variation is still a total
derivative, as described in  \cite{Domokos:2013xqa}.} For
deformations that vanish sufficiently fast at infinity, or at least
leave the boundary conditions unaffected, the first order variation
just vanishes. This is simply the statement that the static
configuration is an {\em extremum} of the potential energy
(including spatial derivatives).

\subsection{Deformed configurations to second order}

The first-order variation condition guarantees that the solution
lies at an extremum of the energy. We can, however, expand the
energy to second order in the small deformation parameter $\xi^i$.
The deformed energy takes the generic form
\begin{align}
E[\phi^a_\Lambda]\approx E[\phi^a_0]+\int d^d\tilde{x}
\tilde{\d}_i\xi^j \ \Pi^i_{\ j}+\int d^d\tilde{x}
\frac{1}{2}\tilde{\d}_i\xi^j\tilde{\d}_l\xi^m \ C^{il}_{jm}+\dots~.
\end{align}
Stability thus requires that
\begin{equation}
\int d^d\tilde{x}
\tilde{\d}_i\xi^j\tilde{\d}_l\xi^m \ C^{il}_{jm} \geq 0.
\end{equation}
Note that this condition must be satisfied for {\em arbitrary} deformations. This is guaranteed if
\begin{equation}\label{weakellip}
C^{il}_{jm}\hat{a}_i \hat{a}_l \hat{b}^j \hat{b}^m \geq 0
\end{equation}
for arbitrary unit vectors $\hat{a}$, and $\hat{b}$.  Flat
directions are those along which the inequality is saturated.

Note that the second order condition (\ref{weakellip}) applies {\em
locally}: unless a solution obeys (\ref{weakellip}), one could
engineer some arbitrarily complicated $\xi^i(x^j)$ in such a way as
to make the integrand negative. Furthermore, the condition holds
even for field configurations of infinite energy and infinite extent
(as long as one can regularize them locally).

This analysis has strong parallels with the theory of elasticity of
solids (see e.g. \cite{Landau}). One can describe the deformation of
a solid in terms of a vector $u^i(x)$, which parametrizes the
displacement of an element of the solid from the equilibrium
configuration at any given point in spacetime. The energy changes as
\begin{equation}
d\varepsilon= \sigma^{ij}u_{ij},
\end{equation}
where $\sigma^{ij}$ is the ``stress'' tensor and $u_{ij}$ is the
``strain'' tensor,
\begin{equation}
u_{ij}=\frac12 (\partial_i u_j +\partial_j u_i)~.
\end{equation}
For instance, the stress tensor of an ideal isotropic fluid is
simply proportional to the pressure, $\sigma^{ij}=-p\delta^{ij}$,
and measures how the energy changes with changes in volume. The
stress tensor can be defined as the variation of the energy with
respect to the strain for adiabatic processes
\begin{equation}
\sigma^{ij}=\left(\frac{\partial \varepsilon}{\partial u_{ij}} \right)_s.
\end{equation}
If the deformations are small ($u_{ij}<<1$), the stress and the
strain are proportional to each other:
\begin{equation}
\sigma^{ij}=S^{ijkl}u_{kl},
\end{equation}
where $S^{ijkl}$ is the ``stiffness'' tensor.
The system is stable if the energy is minimized for all possible deformations. This imposes conditions on the stiffness tensor,
that can be summarized as the strong ellipticity condition: for any two unit vectors $\hat{a}$ and $\hat{b}$
\begin{equation}\label{strongellip}
S^{iljm}\hat{a}_i \hat{a}_l \hat{b}_j \hat{b}_m >0.
\end{equation}
analogous to our equation \ref{weakellip}, if the deformation
$\xi^i$ plays the same role as the displacement vector, and the
tensor $C$ is the stiffness.

\subsubsection{Stiffness in scalar field theories}

Let us now compute the stiffness for a general scalar field theory.
We transform the coordinates as $\tilde{x}^i=x^i-\xi^i(x)$, so the
energy to second order in fluctuations becomes
\begin{align}
E[\phi^a_\Lambda] &=\int d^dx\,\cE\left(\phi^a_\Lambda(x),\partial_i\phi^a_\Lambda(x)\right)\nonumber\\
&=\int d^d \tilde{x} \left|\left|\frac{\partial x^i}{\partial \tilde{x}^j}\right|\right|\cE\left(\phi^a_0(\tilde{x}),\frac{\partial \tilde{x}^j}{\partial x^i}\tilde{\partial}_j \phi^a_0(\tilde{x})\right)\nonumber\\
&= E[\phi^a_0]+\int d^d \tilde{x}\d_i\xi^j\left(\delta^i_j\cE-\tilde{\d}_j\phi^a\frac{\delta\cE}{\delta\tilde{\d_i}\phi^a} \right)\nonumber\\
&\qquad+\int d^d \tilde{x} \ \frac{1}{2}\d_i\xi^j\d_l\xi^m\left[\left(
\delta^i_j\delta^l_m+\delta^i_m\delta^l_j\right)\cE+ \tilde{\d}_j\phi^a\tilde{\d}_m\phi^b\frac{\delta^2 \cE}{\delta\tilde{\d}_i\phi^a\delta\tilde{\d}_l\phi^b}
\right]
\end{align}
where we have used the expansions
\begin{align}
\frac{\partial \tilde{x}^j}{\partial x^i}&=\delta_i^{\
j}-\partial_i\xi^j\\
\left|\left|\frac{\partial x^i}{\partial
\tilde{x}^j}\right|\right|&= 1+\partial_i\xi^i+\frac{1}{2}\partial_i
\xi^l\partial_l \xi^i+\frac{1}{2}(\partial_i\xi^i)^2+O(\xi^3)~.
\end{align}
Rewriting all factors of $\d/\d x^i$ in terms of $\d/\d\tilde{x}^i$, we find the variation of the energy:
\begin{align}
&E[\phi^a_\Lambda]-E[\phi^a_0] = \int d^d \tilde{x}\tilde{\d}_i\xi^j\left(\delta^i_j\cE-\tilde{\d}_j\phi^a\frac{\delta\cE}{\delta\tilde{\d_i}\phi^a} \right)\nonumber\\
&\qquad+\int d^d \tilde{x} \ \frac{1}{2}\tilde{\d}_i\xi^j\tilde{\d}_l\xi^m\left[\left(
\delta^i_j\delta^l_m+\delta^i_m\delta^l_j-2\delta_j^l\delta^i_m\right)\cE+\tilde{\d}_j\phi^a\tilde{\d}_m\phi^b\frac{\delta^2\cE}{\delta\tilde{\d}_i\phi^a\delta\tilde{\d}_l\phi^b} \right. \nonumber\\
&\qquad\qquad+\left.\delta^j_l\tilde{\d}_m\phi^a\frac{\delta\cE}{\delta\tilde{\d_i}\phi^a}+\delta^i_m\tilde{\d}_j\phi^a\frac{\delta\cE}{\delta\tilde{\d_l}\phi^a}-\delta^m_l\tilde{\d}_j\phi^a\frac{\delta\cE}{\delta\tilde{\d_i}\phi^a}-\delta^i_j\tilde{\d}_l\phi^a\frac{\delta\cE}{\delta\tilde{\d_m}\phi^a}
\right].
\end{align}
As noted above, the first order term is a total derivative, so to
leading order the variation gives
\begin{equation}
E[\phi^a_\Lambda]-E[\phi^a_0] =\frac{1}{2}\int d^d \tilde{x}
\;C^{il}_{jm}\tilde{\d}_i\xi^j\tilde{\d}_l\xi^m,
\end{equation}
where the tensor $C$ is
\begin{align}
C^{il}_{jm}&=\left(
\delta^i_j\delta^l_m-\delta_j^l\delta^i_m\right)\cE+\tilde{\d}_j\phi^a\tilde{\d}_m\phi^b\frac{\delta^2\cE}{\delta\tilde{\d}_i\phi^a\delta\tilde{\d}_l\phi^b} \nonumber\\
&\qquad\qquad+\delta_j^l\tilde{\d}_m\phi^a\frac{\delta\cE}{\delta\tilde{\d_i}\phi^a}+\delta^i_m\tilde{\d}_j\phi^a\frac{\delta\cE}{\delta\tilde{\d_l}\phi^a}-\delta_m^l\tilde{\d}_j\phi^a\frac{\delta\cE}{\delta\tilde{\d_i}\phi^a}-\delta^i_j\tilde{\d}_l\phi^a\frac{\delta\cE}{\delta\tilde{\d_m}\phi^a}~.
\end{align}

\subsubsection{Stiffness in gauge field theories}

For theories with vector fields, the definition of $C^{il}_{mn}$ is
slightly more complicated. In some cases it may be more
straightforward to compute the tensor $\tC^{\mu\nu}_{\rho\sigma}$,
whose spatial part is $C^{ij}_{lm}$. We present the computation of
$\tC^{\mu\nu}_{\rho\sigma}$ for gauge-invariant energy densities in
Appendix \ref{app:Ctilde}. This tensor also arises as the quadratic
part of an effective Hamiltonian for time-dependent fluctuations
around the spatially modulated vacuum, an investigation which we
leave for future work.

We now turn to calculating $C^{ij}_{lm}$ for gauge theories (including those with Chern-Simons terms). After the
deformation $x^i\rightarrow \tilde{x}^i$, the energy density depends
on the
\begin{equation}
A_0(\tilde{x}), \ \ \frac{\partial \tilde{x}^k}{\partial x^i}A_k(\tilde{x}), \ \ \frac{\partial \tilde{x}^k}{\partial x^i}F_{0k}(\tilde{x}),\ \  \frac{\partial \tilde{x}^k}{\partial x^i}\frac{\partial \tilde{x}^l}{\partial x^j}F_{kl}(\tilde{x}),
\end{equation}
The expansion of the energy density to second order is now
\begin{align}
\notag &\cE_\Lambda=\cE-\partial_i \xi^l\pi^i_{\ l}\\
&+\frac{1}{2}\partial_i \xi^j\partial_l
\xi^m\left[\delta^l_j\pi^i_{\ m}+\delta^i_m\pi^l_{\
j}-\delta^i_j\pi^l_{\
m}-\delta^l_m\pi^i_{\
j}+2F_{jm}\frac{\delta\cE}{\delta F_{il}}+\phi^{il}_{~~jm} \right]
\end{align}
where we define
\begin{align}
\pi^i_{\ l}&=\frac{\delta \cE}{\delta A_i}A_l+2\frac{\delta
\cE}{\delta F_{i\mu}}F_{l\mu}\\
\phi^{il}_{~~jm}&=\frac{\delta^2\cE}{\delta A_i\delta
A_l}A_mA_j+2\left(\frac{\delta^2\cE}{\delta F_{i\mu}\delta
A_l}A_mF_{j\mu}+\frac{\delta^2\cE}{\delta F_{l\mu}\delta
A_i}A_jF_{m\mu}\right)+4\frac{\delta^2\cE}{\delta F_{i\mu}\delta
F_{l\nu}}F_{j\mu}F_{m\nu}.
\end{align}
As for the case with scalar fields, the change in the energy to second order is
\begin{equation}
E_\Lambda-E =\frac{1}{2}\int d^d \tilde{x} \;C^{il}_{jm}\tilde{\d}_i\xi^j\tilde{\d}_l\xi^m.
\end{equation}
Where the tensor $C$ is
\begin{align}\label{Cgauge}
 &C^{il}_{jm}=\left(
\delta^i_j\delta^l_m-\delta^i_m\delta^l_j\right)\cE+\delta^l_j\pi^i_{\
m}+\delta^i_m\pi^l_{ \ j} -\delta^i_j\pi^l_{\
m}-\delta^l_m\pi^i_{\
j}+2F_{jm}\frac{\delta\cE}{\delta
F_{il}}+\phi^{il}_{~~jm}.
\end{align}
We can new use the result of this and the previous subsection to
analyze the stability of static solutions.

\subsubsection{Global deformations to second order}

In systems with global (or internal) symmetries, a larger set of constraints must be satisfied to guarantee that  the energy is minimized.
Since this analysis is not directly related to the spatially modulated configurations we study in the next section, a reader interested mainly in the latter can proceed directly to section \S \ref{sec:stress}.

As described in \cite{Domokos:2013xqa}, in addition to geometrical
deformations of static solutions, one can also deform static solutions
using global symmetries. This corresponds to elevating the constant
parameters of the global symmetry transformations to space-dependent
ones.
\be \delta_\theta
\phi^a(x) =  \theta^A ({T_A})^b_a \phi_b(x) \quad \Rightarrow
\qquad\delta_\theta \phi^a(x) = \theta^A(x) {T_A}^b_a \phi_b(x)\ .
\ee
Requiring that the energy be extremized under such deformations
yields the constraint
\be \delta_\theta E[\phi^a] =\int d^d x [
\pa_i \theta^A J^i_A + \pa_i (\delta_\theta \Psi^{i0}_0)] \quad
\Rightarrow \qquad \int d^d x J^i_A =0~,
\ee
where $J^i_A$ is a
space component of the current associated with some global symmetry
that is broken by the static solution. $\theta^A$ is the parameter
of the transformation, and in the last step we assumed that the
surface term vanishes and we took $\theta^A= \lambda^A_i x^i$ for constant
$\lambda^A_i$. As above, we can derive an additional condition by
requiring that the extremum be a minimum. Let us expand the energy
to second order in the global deformation. For a constant parameter,
the transformation is a symmetry of the energy. Overall,
\begin{align}\label{symtran}
&0= E[\phi^a_\theta]-E[\phi^a_0]\CR &=\frac12 \int\left [ d^dx \frac{\delta^2 \cE}{\delta \phi^a\delta \phi^b}\delta \phi^a\delta \phi^b + 2 \frac{\delta^2 \cE}{\delta \phi^a\delta \pa_i \phi^b}\delta \phi^a\pa_i(\delta \phi^b) +\frac{\delta^2 \cE}{\delta\pa_i \phi^a\delta \pa_j \phi^b}\pa_i(\delta \phi^a)\pa_j(\delta \phi^b)\right ]\CR
&=\frac12 \int\left [ d^dx\left (\theta^A\theta^B (T_A)^a_c
(T_B)^b_d\right )\times\right.\CR
&\left. \left[  \frac{\delta^2 \cE}{\delta \phi^a\delta
\phi^b}\phi^c\phi^d + 2 \frac{\delta^2 \cE}{\delta\phi^a\delta \pa_i
\phi^b}\phi^c\pa_i\phi^d +\frac{\delta^2 \cE}{\delta\pa_i
\phi^a\delta \pa_j \phi^b}\pa_i(\phi^c)\pa_j( \phi^d)\right ]\right
]\, .\CR
\end{align}
Now we allow the parameters of transformation
$\theta^A$ to be space-dependent. Using  the symmetry transformation
(\ref{symtran}) the variation of the energy is
\bea & &
E[\phi^a_\theta]-E[\phi^a_0] =\frac12 \int\left [ d^dx\left (
(T_A)^a_c (T_B)^b_d  \right )\left[ \frac{\delta^2 \cE}{\delta\pa_i
\phi^a\delta \pa_j
\phi^b}\phi^c\phi^d\right]\pa_i\theta^A\pa_j\theta^B\right]\CR
 &+&\frac12 \int\left [ d^dx\left ( (T_A)^a_c (T_B)^b_d  \right )\left[ \frac{\delta^2 \cE}{\delta\pa_i\phi^a\delta  \phi^b}
  \phi^c\pa_j\phi^d  +\frac{\delta^2 \cE}{\delta\pa_i \phi^a\delta \pa_j \phi^b} (\pa_j\phi^c)\phi^d  )\right ]\pa_i(\theta^A \theta^B) \right ]\, .\CR
\eea
Thus the generalization of the stiffness tensor for global
deformations (we may call it a `susceptibility' tensor) takes the form
\be\label{globmin} E[\phi^a_\theta]-E[\phi^a_0]= \int
d^d x C^{ij}_{AB}\pa_i\theta^A\pa_j\theta^B + C^{i}_{AB}(
\pa_i\theta^A \theta^B + \theta^A \pa_i\theta^B)>0 \ee where
\bea\label{stiffglob} C^{ij}_{AB} &=& \frac12\frac{\delta^2 \cE }{
\delta ( \pa_i\phi^a)\delta(\pa_j\phi^b)}  (T_A) ^a_c (T_B)^b_d
\phi^c \phi^d \CR C^{i}_{AB} &=& \frac12\left [  \frac{\delta^2 \cE
}{ \delta ( \pa_i\phi^a)\delta(\pa_j\phi^b)}  \phi^c \pa_j \phi^d
+\frac{\delta^2 \cE}{\delta\pa_i\phi^a\delta  \phi^b}
  \phi^c\phi^d  \right ] (T_A) ^a_c (T_B)^b_d\,.
\eea
The inequality at the end of  (\ref{globmin}) is of course the requirement that the extremum is indeed a minimum. Obviously
the energy is invariant under  global transformations, namely,
constant $\theta^A$.

For scalar theories with a flavor symmetry and terms at most quadratic in derivatives the susceptibility tensor takes the form

 \be {\cal L}
= G_{ab}(\phi)\frac12 \pa_\mu \phi^a \pa^\mu \phi^b - V(\phi^a)~,
\ee we find
\begin{align}\label{stiffglobtdsig} &C^{ij}_{AB} 
G_{ab}(\phi) (T_A) ^a_c (T_B)^b_d \phi^c \phi^d~, \CR &C^{i}_{AB}
= \frac12\left [  \eta_{ij} G_{ab}(\phi) \phi^c \pa_j \phi^d   +
\pa_bG_{ae}(\phi) \pa_i \phi^e
  \phi^c\phi^d  \right ] (T_A) ^a_c (T_B)^b_d~.
\end{align}
Note that there are terms that are not  only proportional to the derivative of the parameter $\theta^A$ but to the parameter itself. These appear when the symmetry is non-Abelian. If we perform a transformation $\theta^A$ followed by a transformation $\theta^B$, this should be equivalent to a transformation $\theta^C$, but the relation between $\theta^C$, $\theta^A$ and $\theta^B$ is in general non-linear.

We can further extend the analysis by considering both global and geometric deformations together, with possible mixed terms between the two. We will not study this possibility here.

\section{Stress forces}\label{sec:stress}

If a
body is in mechanical equilibrium, the sum of all the forces inside
a volume element vanish. The forces acting on the volume should
be precisely equal to the force of the volume acting on the
surrounding medium.

The force acting on a volume $V$ is the time variation of the momentum
\begin{equation}
F_i=\partial_0 P_i=\int_V d^d x\, \partial_0 T^0_{\ i},
\end{equation}
where $d$ is the number of spatial dimensions and $T^0_{\ i}$ is the
momentum density. In a field theory this is part of the
energy-momentum tensor $T^\mu_{\ \nu}$. Using energy-momentum
conservation we can write the force in terms of the stress tensor as
\begin{equation}
F_i= -\int_V d^d x\, \partial_k T^k_{\ i}=- \oint d^{d-1}\sigma
\hat{n}_k T^k_{\ i}~.
\end{equation}
We define $\hat{n}_k$ as the unit vector orthogonal to the surface,
pointing out from the volume. Therefore, in the absence of external
sources, the force acting on a volume element is determined by the
stresses at the boundary. Consider for example an isotropic medium
with a pressure\footnote{The pressure $p$ could also be negative in
which case we call it a {\em tension}.} that depends on one of the
spatial coordinates $z$:
\begin{equation}
T^k_{\ i}=p(z)\delta^k_{\ i}.
\end{equation}
Focus on a cylindrical block (length $L$, area $A$) within the
material, with its axis along the $z$ direction of length $L$ and
caps of area $A$. The z-direction force acting on the block
\begin{equation}
F_z=-A(p(L/2)-p(-L/2)).
\end{equation}
If $p(L/2)> p(-L/2)$, the cylinder is pushed towards negative values
of $z$ by the higher pressure at $z=L/2$.

We can now apply this type of analysis to solutions periodic along
one spatial direction. We can perform arbitrary local deformations
on the solution -- each such deformation will change the balance of
internal forces in the material. For instance, we can decrease the
size of a unit cell, while at the same time deform the neighboring
cells so the periodic solution remains unchanged farther away. The
forces on the faces of the unit cell no longer cancel: the net force
on the surface after the transformation could be pointing either in
of or out of the unit cell. The first case is clearly unstable,
since the deformed cell will now continue decreasing in size. In the
second case the force tends to push the configuration back toward
the original equilibrium (see figure \ref{fig:stretch}). If the
period of the unit cell is $L=2\pi/k$,  the condition of stability
becomes
\begin{equation}\label{stabilityp}
\partial_k p > 0.
\end{equation}
Simply put, when the cell is compressed its pressure should
increase, and viceversa.

We can realize the situation above by slightly modifying a solution
characterized by a given wavenumber $k$: we make one strip shorter
by a small amount and the neighboring strips larger so that the
solution is not modified further in the material outside the strip.
We can approximate this change by gluing a solution with $k+\delta
k$ between $z=\pm \pi/(k+\delta k)$ and the solution with
$k'=k\frac{k+\delta k}{k+2\delta k}$ in the intervals
$\left[-\frac{3\pi}{k},-\frac{\pi}{k+\delta k} \right]$,
$\left[\frac{\pi}{k+\delta k},\frac{3\pi}{k}\right]$. The solution
for $|z|>3\pi/k$ is the same as before.

\FIGURE[t!]{ \centering
\includegraphics[width=3.5in]{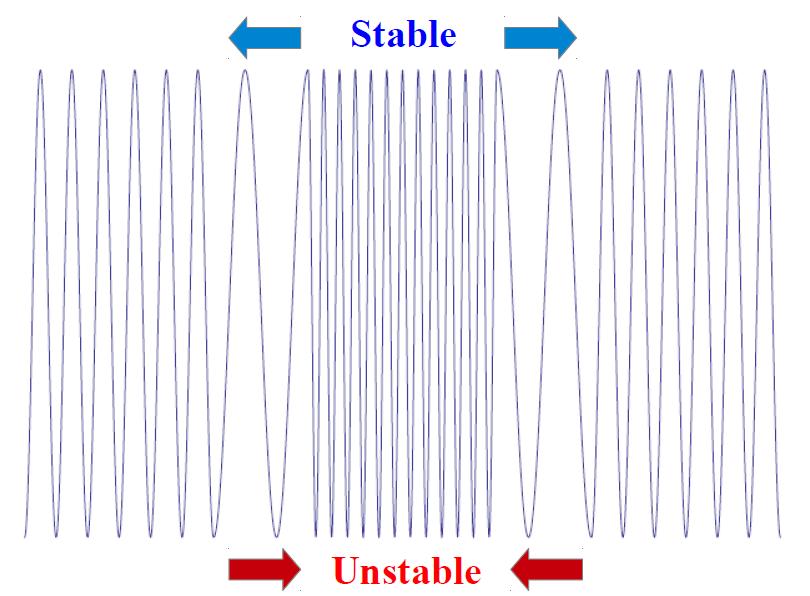}
\caption{When a cell is compressed relative to its neighboring cells there are two possible situations. In the first case (upper arrows in blue), the force on the surfaces of the compressed cell point outward, so the force is restoring and the cell will go back to its original size when external forces are turned off. In the second case (lower arrows in red), the surface forces point inward, so the cell will continue compressing even further instead of returning to its original size.}
\label{fig:stretch}
}

We can also perform other deformations, such as changing the shape
of the cell. In all cases the system will be stable if forces are
restoring and unstable otherwise.

We apply this argument to spatially modulated solutions in
holographic models of QCD in section \S \ref{sec:spatiallymodulated}.

\section{Application to spatially modulated solutions}\label{sec:spatiallymodulated}

A spatially inhomogeneous vacuum (in holographic
models with  Chern-Simons term) was
identified in \cite{Ooguri:2010kt,Ooguri:2010xs}. \cite{Ooguri:2010kt} studies Maxwell-Chern-Simons (MCS)
theory in an $AdS_5$ black hole, while \cite{Ooguri:2010xs} treats
gauge fields living on D8 branes in the compactified D4 black hole
geometry. The latter is the deconfined phase of the Sakai-Sugimoto
model \cite{Sakai:2004cn,Aharony:2006da}. Both examples were studied
in the probe approximation, with no backreaction from the metric.
Later works (e.g. \cite{Donos:2012wi}) include backreaction in the
MCS theory, with qualitatively similar results. For simplicity we
will study  the stability conditions in the probe MCS in detail, and
later on comment on the backreacted solutions.

\subsection{Spatially modulated phases in $AdS_5$ black holes}

A simple example of spatially modulated structure exists in
 in Maxwell-Chern-Simons theory \cite{Ooguri:2010kt},
\begin{align}
\cL=\frac{\sqrt{-g}}{\alpha^2}\left[-\frac{1}{4}\tF_{IJ}\tF^{IJ}+\frac{1}{3!\sqrt{-g}}\epsilon^{IJKLM}\tA_I\tF_{JK}\tF_{LM}~,
\right]
\end{align}
in an $AdS_5$ black hole background. The $\tA_I = \alpha A_I$ are
gauge fields rescaled with respect to the Chern-Simons coupling such
that entire action becomes proportional to $\alpha^{-2}$.

The holographic dual of this model is a conformal field theory at
finite temperature with a probe sector having a global symmetry. The
associated global current is dual to the gauge field in the bulk,
and the Chern-Simons term implies that the global symmetry is
anomalous in the field theory. This model is interesting as a
simplified version of QCD at finite temperature, taking into account
effects of the chiral anomaly.

The authors of \cite{Ooguri:2010kt} explore this model in the limit
$\alpha\rightarrow\infty$ with a finite chemical potential.
($A_0(r=\infty)=\mu\rightarrow 0$ with $\alpha\mu$ finite.) The background
metric is given by
\begin{equation}
ds^2=-H(r)dt^2+H(r)^{-1}dr^2+r^2d\vec{x}^2\qquad\text{with}\qquad
\vec{x}=(x_2,x_3,x_4)
\end{equation}
and
\begin{align}
H(r)=r^2\left[1-\left(\frac{r_+}{r}\right)^4 \right]~.
\end{align}
The background electric field corresponds to the asymptotic value of
the field strength,
\begin{align}
\lim\limits_{r\rightarrow\infty}\tF_{0r}=\frac{\tE}{r^3}=-\frac{2r_+^3}{\pi\tau
r^3}\quad
\end{align}
where $\tau=T/\mu\alpha$ is the rescaled temperature of the black
hole. A spatially modulated solution
of the equations of motion is found in \cite{Ooguri:2010kt}, of the form
\begin{align}\label{eq:OPansatzYMCS}
\tA_0=f(r), \quad \tA_3+i\tA_4=h(r)e^{ikx_2}
\end{align}
with all other gauge field components vanishing.\footnote{We have
absorbed a negative sign into $k$ with respect to the convention of
\cite{Ooguri:2010kt} for later convenience.} The equations of motion
under this ansatz are
\begin{align}
\d_r\left(r^3f'+2kh^2\right)&=0 \\ \d_r\left(r
Hh'\right)-\frac{k^2}{r}h+4f'kh&=0~.
\end{align}
Integrating the first line allows us to express $f'$ with an
integral of motion $\tE$, using
\begin{align}
r^3f'+2kh^2=-\tE~.
\end{align}
This leaves a single equation
\begin{align}
r^3\d_r\left(rHh' \right)-r^2k^2h-4kh\left(\tE+2kh^2 \right)=0~.
\end{align}
As explained in \cite{Ooguri:2010kt}, this equation has non-trivial
solutions -- that is, with amplitude $h(r_+)=h_0$ nonzero -- only
for a limited set of values of $k$. Solving the above equation under the
boundary condition and assuming no sources for the currents (i.e. no
non-normalizable modes in $A_3,A_4$) one finds a relation between $h_0$
and the wavenumber of the modulation, $k$ (see
 Figure \ref{fig:AdSbhh0k}).

\FIGURE[t!]{
\includegraphics[width=8.5cm]{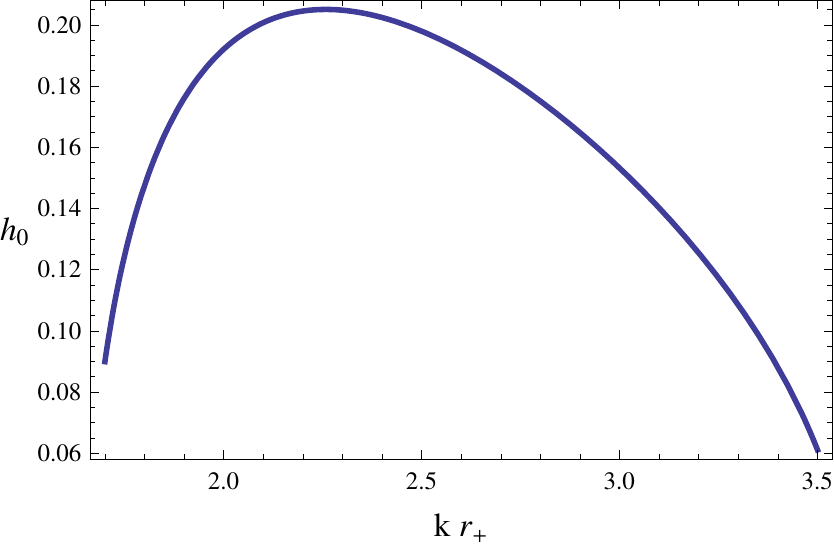}
    \caption{Amplitude $h_0$ of the spatially modulated solution as a function of wave number, $k$ in units of
$1/r_+$. We take $\tau=0.35$.}
     \label{fig:AdSbhh0k}
}

In holographic models with D-branes the components of the
energy-momentum tensor of the brane integrated along the radial
direction are identified with the energy-momentum tensor of the dual
field theory \cite{Karch:2008uy},
\begin{equation}
\vev{T^\mu_{\ \nu}}=\int_{Dp} dr \sqrt{g_{Dp}} T^\mu_{Dp\,\ \nu},
\end{equation}
as a result of the map between symmetries of the gravity and gauge
theories. It is natural to extend this map to the probe gauge
fields, so that the energy density of the field theory can be
defined as the integral of the bulk $T_{00}$ along the radial
direction
\begin{equation}
\vev{\cE}=\int_{r_+}^\infty dr \left[\frac{1}{2r^3}\left(
\tE+2kh^2\right)^2+\frac{k^2h^2}{2r}+\frac{rHh'^2}{2}\right]~.
\end{equation}
This is also identified with the free energy,\footnote{There is, in
addition, a boundary contribution that we have not included.
However, when we Legendre transform to an ensemble with fixed
density, this term cancels out. The details can be found in the
Appendix of \cite{Bayona:2011ab}.} plotted in Fig.
\ref{fig:AdSBHFE}. Its lies at roughly $kr_+=2.38$.\footnote{The
energy density, equation 3.11 in \cite{Ooguri:2010kt} seems to be
missing a $r^3$ factor in the integrand.} The corresponding solution
is thus thermodynamically stable within this family of spatially
modulated solutions.
\FIGURE[t!]{
\includegraphics[width=8.5cm]{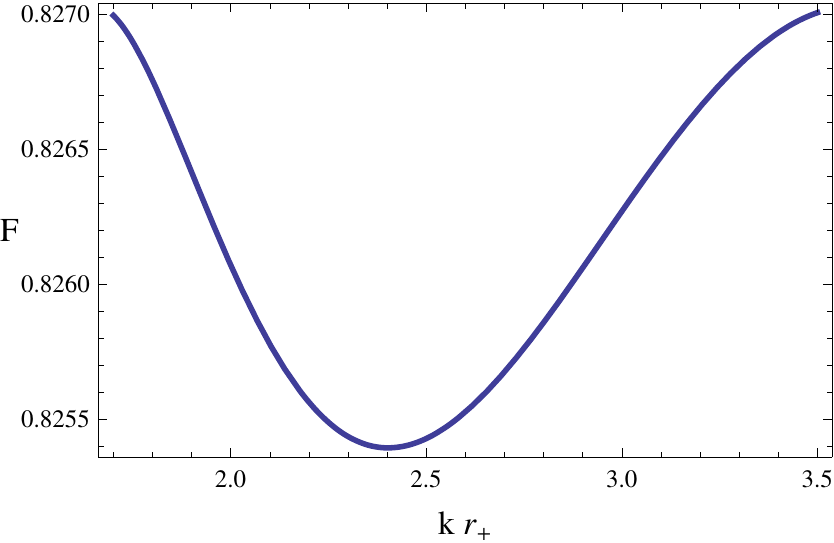}
    \caption{Free energy $F$ of the spatially modulated solution as a function of wave number, $k$, in units of
$1/r_+$. We set $\tau=0.35$. The function is minimized for
$kr_+\simeq 2.4$.}
     \label{fig:AdSBHFE}
}

\subsubsection{Stiffness tensor}

Using the results of section \S~\ref{sec:elastic} we can compute the
stiffness tensor to check
the stability of these solutions under small geometrical deformations. The energy
density is
\begin{equation}
\cE=\int_{r_+}^\infty dr\, \sqrt{-g}\left[\frac{1}{2}|g^{00}|g^{ij} F_{0i} F_{0j}+\frac{1}{2}|g^{00}|g^{rr} F_{0r} F_{0r}+\frac{1}{2}g^{rr}g^{ij} F_{ri} F_{rj}+\frac{1}{4}g^{ik}g^{jl}F_{ij} F_{kl}\right].
\end{equation}
The expression for the stiffness tensor is quite complicated, but one can show that the condition \eqref{weakellip} is satisfied.
We illustrate only a part of the result. Let us take the unit vectors
\begin{equation}
\hat{a}=(\cos\theta_1,\sin\theta_1\cos \varphi_1,\sin\theta_1\sin \varphi_1), \ \ \hat{b}=(\cos\theta_2,\sin\theta_2\cos \varphi_2,\sin\theta_2\sin \varphi_2).
\end{equation}
Then,
\begin{align}
\notag C^{il}_{jm}\hat{a}_i \hat{a}_l \hat{b}^j \hat{b}^m &=\frac{1}{32}\int_{r_+}^\infty dr\,\frac{k^2h^2}{r}\left[18 -\cos \left(2 k x_2+4 \theta _1-2 \phi _1\right)-\cos \left(2 k x_2-2 \left(2 \theta _1+\phi
   _1\right)\right)\right.\\
\notag   &   +64 \sin \left(\theta _1\right) \sin \left(\theta _2\right) \cos ^2\left(\theta _1\right) \sin \left(k x_2-\phi
   _1\right) \sin \left(k x_2-\phi _2\right)\\
\notag   & -8 \sin ^2\left(\theta _1\right) \cos \left(2 k x_2-2 \phi _2\right)+\cos \left(2
   \theta _2\right) \left(8 \sin ^2\left(\theta _1\right) \cos \left(2 k x_2-2 \phi _2\right)-4\right)\\
\notag   &  +2 \cos \left(2 k x_2-2 \phi
   _1\right)+12 \cos \left(2 \theta _1\right)+2 \cos \left(4 \theta _1\right)+2 \cos \left(2 \left(\theta _1-\theta
   _2\right)\right)\\
   & \left. +2 \cos \left(2 \left(\theta _1+\theta _2\right)\right)\right].
   \label{AdSstiffness}
\end{align}
Note that there is a shift symmetry
\begin{equation}
x_2\to x_2+\delta x_2, \ \ \phi_1\to \phi_1+k\delta x_2, \ \ \phi_2\to \phi_2+k\delta x_2,
\end{equation}
which allows us to evaluate the tensor at a fixed value of $x_2$, for instance $x_2=0$, this gives an expression of the form
\begin{align}
C^{il}_{jm}\hat{a}_i \hat{a}_l \hat{b}^j \hat{b}^m &=\frac{1}{32}\int_{r_+}^\infty dr\,\frac{k^2h^2}{r}\cC(\theta_1,\theta_2,\phi_1,\phi_2).
   \label{AdSstiffness2}
\end{align}
One can easily check that there are some flat directions where $\cC=0$, for instance for $\theta_1=\pi/2$, $\theta_2=0$.
We have also checked that \eqref{AdSstiffness2} is never negative,
by first finding the values of $(\theta_1,\theta_2,\phi_1,\phi_2)$ where $\cC$ is extremized
\begin{equation}
\partial_A\cC=0, \ \ A=\theta_1,\theta_2,\phi_1,\phi_2,
\end{equation}
and evaluating $\cC$ at those points. For all extremal points
$\cC\geq 0$, so \eqref{weakellip} is satisfied and there are no
unstable directions. Note that this result is valid for any
spatially modulated solution on the curve in Figure \ref{fig:AdSbhh0k},  not only for the one which minimizes the free
energy.

\subsubsection{Adiabatic deformations}

In addition to obtaining information about the stability of the
solutions under geometric deformations, we can compute the work done by adiabatic changes.  We introduce
time-dependent deformations and eventually neglect contributions
from time derivatives.

We take $\xi^0=\xi^r=0$ for the deformations. Note  that the deformed configurations are still normalizable, since to second order
\begin{equation}
A_M\to A_M-\partial_M\xi^N A_N+\partial_M \xi^L\partial_L\xi^N A_N.
\end{equation}
The only possible non-normalizable contribution would appear from a term proportional to $A_0$, but since $\xi^0=0$, no such term exists.

The first order term in the variation of the energy is a total derivative, which we take to vanish, plus an additional term proportional to
\begin{equation}
-2\, \partial_0\xi^i\frac{\delta \cE}{\delta F_{0\alpha}}F_{i\alpha}.
\end{equation}
This term appears because there is a net flow of momentum in the
spatially modulated solution. $\partial_0 \xi^i$ is the velocity of
the `volume element' and the energy increases or decreases depending
on whether the volume element moves in the same direction as the
momentum flow. In the spatially modulated background the only
non-zero terms appear for $\alpha=r$ and $i=3,4$, and are
proportional to
\begin{equation}
F_{3r}=-\partial_r A_3=-h' \cos(k x_2), \ \ F_{4r}=-\partial_r A_4=-h' \sin(k x_2).
\end{equation}
We see from this expression that momentum flows in the $x_3$, $x_4$
plane and that its direction rotates as we move along the $x_2$ direction. The
first order term will vanish for deformations transverse to the momentum flow, namely
\begin{equation}\label{eq:timefluct}
\xi^2=\xi^L(t,x,r), \ \ \xi^3=\xi^T(t,x,r)\sin(k x_2), \ \
\xi^4=-\xi^T(t,x,r)\cos(k x_2).
\end{equation}
We compute the quadratic order contributions to the energy
for these deformations, using the results of
Appendix~\ref{app:Ctilde}. From equation (\ref{eq:timefluct}) we can
see that $\tC^{\alpha\gamma}_{0\delta}$, $\tC^{\alpha\gamma}_{\beta
0}$,$\tC^{\alpha\gamma}_{r\delta}$ and $\tC^{\alpha\gamma}_{\beta
r}$ are irrelevant to us (we assume $\xi^0=\xi^r=0$).

The final result for the change in the energy density is
\begin{align}
\cE_2=\cE_T+\cE_L+\cE_{\rm int},
\end{align}
where ($a,b=3,4$)
\begin{align}
\cE_L &=r k^2 h^2\left(\frac{1}{H}(\d_t\xi^L)^2+H(\d_r\xi^L)^2+\frac{1}{r^2}(\d_2\xi^L)^2+\frac{1}{r^2}L^{ab}\partial_a\xi^L\partial_b \xi^L \right),\\
\cE_T &=r k^2 h^2\left(\frac{1}{H}(\d_t\xi^T)^2+H(\d_r\xi^T)^2+H\partial_r\left(hh'(\xi^T)^2\right)+\frac{1}{r^2}\delta^{ab}\partial_a\xi^T\partial_b\xi^T\right),\\
\cE_{\rm int} &= 2k f' h' \xi^T\partial_t\xi^L+\partial_2\xi^L M_{LT}^a\partial_a\xi^T+\partial_2\xi^T M_{TL}^a\partial_a\xi^L+\xi^T N^a\partial_a\xi^L.
\end{align}
We have defined
\begin{align}
L^{ab} &= \hat{z}^a \hat{z}^b, \\
M_{LT}^a &=\left(\frac{k^2 h^2}{r}-r^3(f')^2-rH(h')^2\right)\hat{y}^a,\\
M_{TL}^a &=\left(-3\frac{k^2 h^2}{r}+r^3(f')^2+rH(h')^2\right)\hat{y}^a,\\
N^a &=\left(\frac{k^2 h^2}{r}-r^3(f')^2+rH(h')^2\right)\hat{z}^a,
\end{align}
where the unit vectors $\hat{y}$ and $\hat{z}$ are
\begin{equation}
\hat{y}=(-\sin(k x_2),\, \cos(k x_2)), \ \ \hat{z}=(\cos(k x_2),\, \sin(k x_2)).
\end{equation}
Note that $\hat{z}$ is the direction of the momentum flow in the spatially modulated solution and $\hat{y}$
the direction orthogonal to it in the plane transverse to $x_2$.
Let us now define the coordinates
\begin{equation}
y=\hat{y}^a x_a, \ \ z=\hat{z}^a x_a.
\end{equation}
Note that for $x_2=0$, $y=x_4$ and $z=x_3$. For $x_2\neq 0$, the $y,z$ coordinates are related to the $x_3, x_4$ coordinates by a rotation.

We now assume that $\xi^T=\xi^T(t,r,y,z)$, $\xi^L=\xi^L(t,r,y,z)$. For a function $F(y,z)$ the spatial derivatives take the form ($A,B=y,z$)
\begin{align}
\partial_a F &=\hat{y}_a \partial_y F+\hat{z}_a\partial_z F,\\
\partial_2 F &=y \partial_z F-z\partial_y F\equiv \epsilon^{AB}x_A\partial_B F.
\end{align}
This allows us to simplify the energy density in such a way that the dependence on the coordinates is polynomial:
\begin{align}
\cE_L &=r k^2 h^2\left(\frac{1}{H}(\d_t\xi^L)^2+H(\d_r\xi^L)^2+\frac{1}{r^2}(\epsilon^{AB}x_A\partial_B\xi^L)^2+\frac{1}{r^2}(\d_z \xi^L)^2 \right),\\
\cE_T &=r k^2 h^2\left(\frac{1}{H}(\d_t\xi^T)^2+H(\d_r\xi^T)^2+\frac{H}{h^2}\partial_r\left(hh'(\xi^T)^2\right)+\frac{1}{r^2}(\partial_y\xi^T)^2+\frac{1}{r^2}(\partial_z\xi^T)^2\right),\\
\notag \cE_{\rm int} &= 2k f' h' \xi^T\partial_t\xi^L+(M_{LT}^i\hat{y}_i)\epsilon^{AB}x_A\partial_B\xi^L \partial_y\xi^T+( M_{TL}^a\hat{y}_a)\epsilon^{AB}x_A\partial_B\xi^T \partial_y\xi^L\\ &+(N^a\hat{z}_a)\xi^T \partial_z\xi^L.
\end{align}
We can further simplify these expressions if we take $\xi^T=c\xi^L=c\xi$ for some constant $c$. Then,
\begin{align}
\notag\cE_2 &=(1+c^2)r k^2 h^2\left(\frac{1}{H}(\d_t\xi)^2+H(\d_r\xi)^2+\frac{1}{r^2}\frac{(\epsilon^{AB}x_A\partial_B\xi-c\partial_y\xi)^2}{1+c^2}+\frac{1}{r^2}(\d_z \xi)^2 \right)\\
&+ ck f' h' \partial_t(\xi^2)+\frac{c}{2}(N^a\hat{z}_a) \partial_z(\xi^2).
\end{align}
Note that
\begin{equation}
\epsilon^{AB}x_A\partial_B\xi-c\partial_y\xi=y\partial_z\xi-(z+c)\partial_y\xi,
\end{equation}
so if we define a new coordinate $u=z+c$, the energy density is simply
\begin{align}
\notag\cE_2 &=(1+c^2)r k^2 h^2\left(\frac{1}{H}(\d_t\xi)^2+H(\d_r\xi)^2+\frac{1}{r^2}\frac{((y\d_u-u\d_y)\xi)^2}{1+c^2}+\frac{1}{r^2}(\d_u \xi)^2 \right)\\
&+c k f' h' \partial_t(\xi^2)+\frac{c}{2}(N^a\hat{z}_a) \d_z(\xi^2).
\end{align}
We can read some interesting physics from this expression. First, when we integrate the term $\propto \partial_t(\xi^2)$ over time, it gives a change in the energy proportional to the square of the total displacement. It is analogous to a spring with a spring constant
\begin{equation}
K_{\rm spring}=c \int_{r_+}^\infty  dr\, k f' h'.
\end{equation}
If we set $c=0$ -- that is, if the deformation is only a displacement along the $x_2$ direction -- then $K_{\rm spring}=0$ and the variation of the energy will
 depend only on derivatives of the displacement field if we make $\partial_r\xi=0$. Note that beyond the adiabatic regime this is an issue, since the term
 proportional to $(\partial_t\xi)^2$ diverges at the horizon. Neglecting the time derivatives we are left with the contribution from the spatial components of the stiffness tensor
\begin{align}
\notag\cE &= \frac{k^2 h^2}{r^2}\left(((y\d_z-z\d_y)\xi)^2+(\d_z \xi)^2 \right).
\end{align}
The first term is the square of the angular momentum in the $y, z$ plane, while the
second term is the square of the linear momentum in the $z$ direction. Recall that $z$ is the direction of the flow of momentum in the spatially modulated solution
and that the $y, z$ plane rotates relative to the $x_3, x_4$ axes along the $x_2$ direction. This result suggests the possibility of having modes with unusual dispersion relations, which depend on the angular momentum in the transverse plane rather than on the usual linear momentum.

\subsubsection{Stress forces}

Let us now check the stability condition \eqref{stabilityp} derived from the
condition on the stress forces when the periodicity of the solution is changed locally. The pressure along the direction of spatial modulation is
\begin{align}
p_2=\sqrt{-g}\,T^2_{\phantom{2}2}=-\frac{1}{2r^3}\left(
\tE+2kh^2\right)^2-\frac{k^2h^2}{2r}+\frac{rHh'^2}{2}~.
\end{align}
Using the relation between the boundary and bulk stress tensor, we
identify the pressure in the field theory dual as
\begin{equation}
\vev{p_2}=\int_{r_+}^\infty dr\,\sqrt{-g}\, T^2_{\phantom{2}2}.
\end{equation}
The pressure is negative (it is actually a tension) for all $k$ and
has a maximum at around $k r_+ \simeq 2.25$ (see Fig.
\ref{fig:AdSBHp}). Note that in the region where $\partial_k
\vev{p_2}<0$ a deformation that changes the value of $k$ in a finite
region will {\em not} go back to the original equilibrium. The
minimum of the free energy is inside this region, at $k r_+\simeq
2.4$. This suggests that either the endpoint of the instability of
the homogeneous solution is a different kind of inhomogeneous
solution, or that the probe approximation does not suffice to
describe the energetics of the configuration. We will comment more
on this in the next section.

\FIGURE[t!]{
\includegraphics[width=8.5cm]{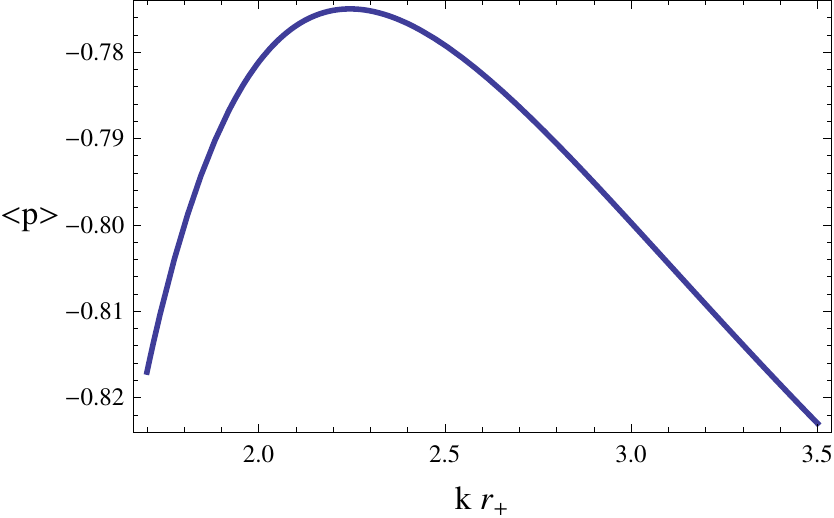}
    \caption{Average pressure of the spatially modulated solution as a function of wave number, $k$, in units of
$1/r_+$. We set $\tau=0.35$. The pressure has a maximum at
$kr_+\simeq 2.25$.}
     \label{fig:AdSBHp}
}

Our arguments so far were based on deformations of the solutions into
configurations that do not satisfy the equations of motion and
therefore require the presence of external forces. Based on the
analysis of stresses we have argued that after the external forces
are turned off the evolution of the system will take it away from
the original solution if the stress forces are not restoring. In
order to show that such initial states are indeed possible we give
an explicit example of such an initial condition in
Appendix \ref{initialc}.

\subsection{Stress forces and thermodynamic stability in backreacted solutions}

Spatially modulated solutions of Maxwell-Chern-Simons, including
backreaction, were constructed by Donos and Gauntlett in
\cite{Donos:2012wi}. Equilibrium configurations were found
numerically by computing the solutions as a function of the
temperature and $k$ and minimizing the free energy at each
temperature. In a more recent paper \cite{Donos:2013cka}, the
authors derive an analytic condition that determines the
minimum of the free energy with respect to $k$.\footnote{In principle one can derive the same relations using the results of \cite{Papadimitriou:2005ii}. We thank Ioannis Papadimitriou for pointing this out to us.} 

Their derivation is
not limited to MCS, but extends to other spatially modulated
solutions as well. In the spontaneously broken case,
\begin{equation}
0=k\frac{\delta w}{\delta k}\Big|_{eq} =w+\vev{p_2}\Big|_{eq}.
\end{equation}
This implies that
\begin{equation}
k\frac{\delta^2 w}{\delta k^2}\Big|_{eq}=\partial_k\vev{p_2}\Big|_{eq}.
\end{equation}
Therefore, the condition on the pressure we have found using the
stress force analysis \eqref{stabilityp} is equivalent to the
condition that the free energy has a minimum. This suggests that
there may be an issue with the definition of the free energy in the
probe approximation. If the free energy were properly
defined, the equilibrium configuration would be located in the stable
region as determined by the stress force analysis.

\subsection{Spatially modulated solutions in the Sakai-Sugimoto model}
The Sakai-Sugimoto model consists of a
D8 brane embedded in the D4 soliton geometry, localized in the 5th compact direction and wrapping an $S^4$. At low
energies it is the holographic dual to a confining QCD-like theory. As temperature is increased, the system undergoes a
transition to a deconfined phase, described by a black hole geometry. In this phase (for large enough temperatures) the D8 splits in a $D8$ and $\bar{D8}$ falling straight into the horizon. This is seen as the dual version of chiral symmetry restoration. The global chiral symmetries map to gauge fields on the branes worldvolumes,
and the chiral anomalies to five-dimensional Chern-Simons terms.

A D8 brane embedded in the D4 black hole geometry has an effective action of the form
\begin{equation}
S_{D8}=-T_8\int dt d^3x du u^{1/4}\sqrt{-\det(g_{\alpha\beta}+\tF_{\alpha\beta})}+\frac{\alpha}{6}T_8\int dt d^3 x du \epsilon^{\mu_1\mu_2\mu_3\mu_4\mu_5}\tA_{\mu_1}\tF_{\mu_2\mu_3}\tF_{\mu_4\mu_5}.
\end{equation}
We have integrated already over the $S^4$ directions, so the effective metric is five-dimensional
\begin{equation}
ds^2=u^{3/2}(-f(u)dt^2+d\vec{x}^2)+\frac{du^2}{u^{3/2}f(u)}, \ \ f(u)=1-\frac{u_T^3}{u^3}.
\end{equation}
The value of the Chern-Simons coupling is $\alpha=\frac{3}{4}$. The authors of \cite{Ooguri:2010xs} demonstrate that a homogeneous D8 embedding with an electric field $F_{0u}$ is unstable for fluctuations of the gauge field with momentum below some threshold. For a density $\rho=5 u_T^{5/2}$ this threshold lies at $k=2.39 u_T^{1/2}$. In the field theory dual, this implies that a finite density homogeneous state is unstable towards the appearance of inhomogeneous currents.

\cite{Ooguri:2010xs}  propose as the endpoint of this  instability a spatially modulated phase that in the holographic dual is described by a configuration of the D8 gauge fields:
\begin{equation}
\tA_t=a(u), \ \ \tA_x+i \tA_y=h(u) e^{-ikz}.
\end{equation}
The solution proceeds as for the AdS black hole case in the previous subsection. Under the ansatz for the solution, the equations of motion become
\begin{align}
0=& \partial_u\left(\frac{u a'(u) \sqrt{k^2 h(u)+u^3}}{\sqrt{-a'(u)^2+f(u) h'(u)^2+1}}\right)+4 k \alpha  h(u) h'(u),\\
0=&\partial_u\left(\frac{u f(u) \sqrt{k^2 h(u)+u^3} h'(u)}{\sqrt{-a'(u)^2+f(u) h'(u)^2+1}}\right)+4 k \alpha  h(u) a'(u)-\frac{k^2 u h(u) \sqrt{-a'(u)^2+f(u) h'(u)^2+1}}{\sqrt{k^2 h(u)+u^3}}.
\end{align}
The first equation can be integrated, and solved for $a'(u)$:
\begin{equation}
a'(u)=\frac{\sqrt{f(u) h'(u)^2+1} \left(\rho -2 k \alpha  h(u)^2\right)}{\sqrt{k^2 u^2 h(u)+4 k^2 \alpha ^2 h(u)^4-4 k
   \alpha  \rho  h(u)^2+u^5+\rho ^2}}.
\end{equation}
 $\rho$ is an integration constant proportional to the charge density.

The equation of motion for $h(u)$ becomes
\begin{equation}
K(u)\partial_u\left(K(u)f(u)h'(u) \right)+4 k \alpha  h(u) \left(\rho -2 k \alpha  h(u)^2\right)-k^2 u^2 h(u)=0,
\end{equation}
where
\begin{equation}
K(u)=\sqrt{\frac{k h(u)^2 \left(4 k \alpha ^2 h(u)^2+k u^2-4 \alpha  \rho \right)+u^5+\rho ^2}{f(u) h'(u)^2+1}}.
\end{equation}

In order to solve the equation of motion one needs to fix the initial conditions at the horizon $h_0=h(u_T)$ and
\begin{equation}
h'(u_T)=\frac{h_0 k u_T \left(8 h_0^2 k \alpha ^2+k u_T^2-4 \alpha  \rho \right)}{3 \left(h_0 k \left(4 h_0^3 k \alpha
   ^2-4 h_0 \alpha  \rho +k u_T^2\right)+u_T^5+\rho ^2\right)}.
\end{equation}
One can determine the initial condition for the derivative by evaluating the equations of motion at the horizon and eliminating the term $a''(u_T)$.

For the numerical calculation we rescale coordinates, density and momentum in such a way that the horizon is effectively at $u_T=1$. Following \cite{Ooguri:2010xs}, we then use the shooting method to solve for solutions with $\rho=5$, $\alpha=3/4$,  demanding that the solution be normalizable at the boundary.
We find solutions in the interval $k\in [1.485,4.3494]$ that resemble closely those found by Ooguri and Park. The value of the initial condition $h_0$ for different values of $k$ is plotted in figure \ref{fig1}.

\FIGURE[t!]{
\includegraphics[width=8.5cm]{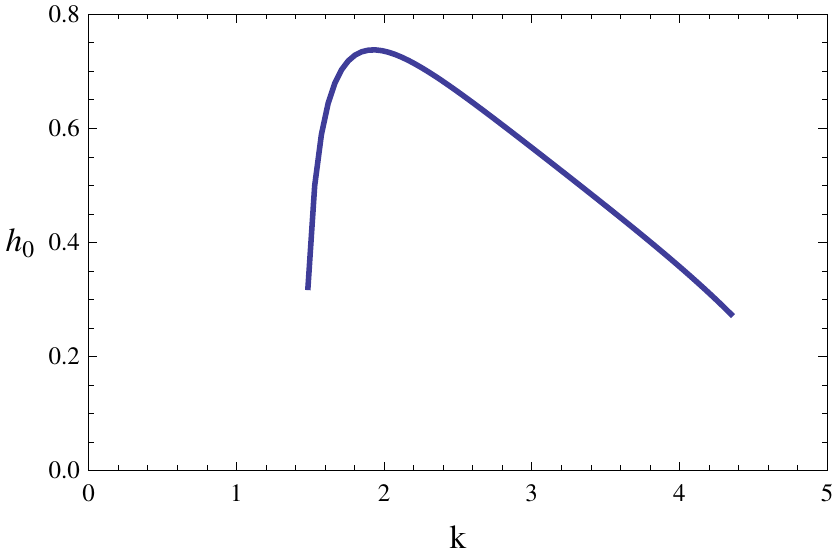}
    \caption{Initial values $h_0$ for different values of $k$.}
     \label{fig1}
}

We can now check whether this solution is stable under the deformation conditions.
The longitudinal pressure density in the $z$ direction is
\begin{equation}
p_z=T^z_{\ z}=2\frac{\partial \cL}{\partial F_{zx}}F_{zx}+2\frac{\partial \cL}{\partial F_{zy}}F_{zy}-\cL.
\end{equation}
This gives
\begin{equation}
p_z=T_8 u^4 \sqrt{\frac{-a'(u)^2+f(u) h'(u)^2+1}{k^2 h(u)^2+u^3}}.
\end{equation}
The pressure density is divergent when $u\to \infty$
\begin{equation}
p_z\simeq u^{5/2}+O\left(u^{-5/2}\right),
\end{equation}
but the divergence is independent of the solution, so we can simply subtract it in order to compute the finite pressure in the dual field theory
\begin{equation}
\vev{p_z}=\int_{u_T}^\infty du\, \left(p_z-u^{5/2}\right).
\end{equation}
The result is plotted in figure \ref{fig2}. We observe that the condition $\partial_k\vev{p_z}>0$ is satisfied only in a region of low $k$, $k\lesssim 1.89 u_T^{1/2}$. As for the case of Maxwell-Chern-Simons in the probe approximation, the stress force analysis seems to be in disagreement with the quoted values of the free energy minimum. We have seen that in the backreacted MCS solution the stress force condition is consistent with the thermodynamic analysis, so for the D8 branes it may also be related to the probe approximation and/or the definition of the free energy.  The MCS example suggests that the issue may be solved by taking into account the backreaction and the contribution of closed string fields to the free energy.

\FIGURE[t!]{
\includegraphics[width=8.5cm]{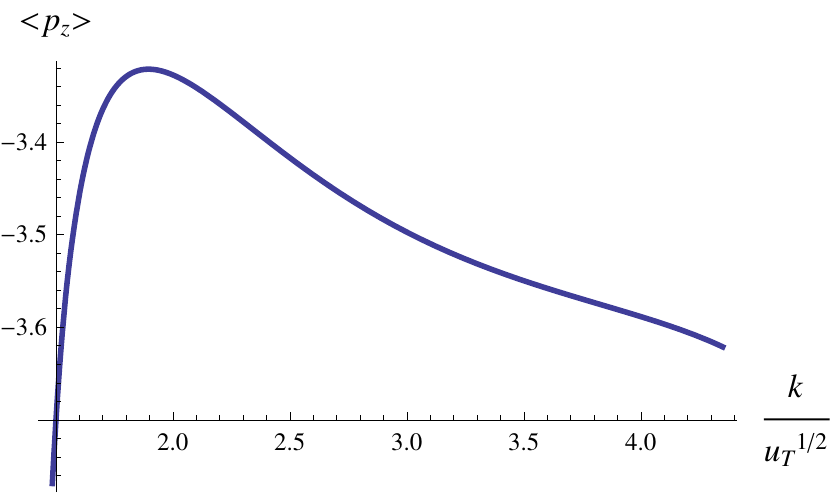}
    \caption{Pressure of the spatially modulated solutions in the Sakai-Sugimoto model. The momentum is normalized in units of $u_T^{1/2}$ and the pressure in units of the effective D8 brane tension. When the slope is positive the solutions are stable according to the stress force analysis.}
     \label{fig2}
}

\section{Conclusions and future directions}\label{sec:end}

We have presented new tests of stability for inhomogeneous phases
based on small geometric deformations. We used these tests to verify
the stability of spatially modulated solutions, which are
holographic duals of finite density states. First, we  expanded the
energy to second order in the deformations and established under what condition the
solution is a local minimum. We then presented a version of these
conditions generalized to global symmetries. The second test of stability amounts to demanding that local deformations
induce a restoring force in the material. For the solutions
constructed in \cite{Ooguri:2010kt} we found that the energy minimization condition is satisfied, but the force condition is
not satisfied. The force condition is also not satisfied in general in \cite{Ooguri:2010xs}, and furthermore the minimum of the free
energy lies outside the stable region.
However,  analyzing the backreacted solution for the system of \cite{Ooguri:2010kt} (using
the results of \cite{Donos:2013cka}), we find that the restoring
force condition and thermodynamic stability are equivalent. This
suggests that the free energy of the probe solution has not been
properly defined. In other words, that simply minimizing the energy
of the brane while neglecting the background is not a valid
approximation. Another possibility is that the probe solution is indeed unstable.

Although we have worked out the details of only a few examples, the
methods we present here can be applied to a variety of inhomogeneous
solutions. An obvious advantage of the stability
checks we have presented is that they do not require solving
the linearized fluctuation equations for time-dependent
configurations. Of course this renders the analysis less
comprehensive, since there may be unstable directions in field space
not captured by the kind of deformations we have presented.

The
stability analysis can be improved by including additional types of
deformations, such as
global (or gauge) symmetry deformations. This extension is relevant to myriad holographic applications, such as solutions involving
non-Abelian gauge fields dual to  p-wave superfluids
\cite{Gubser:2008wv,Ammon:2008fc}.

A further extension of this work would exploit the connection
between Goldstone bosons and symmetry transformations. As the
solutions we study spontaneously break translation and/or rotation
invariance, we expect that at low energies there should be gapless
modes, the Goldstone bosons. The precise number and dispersion
relations of these modes is a more complicated issue when Lorentz
invariance is broken (see for instance
\cite{Nielsen:1975hm,Watanabe:2011ec}, and \cite{Amado:2013xya} for
an example in holography). In principle we expect the Goldstone
modes to take a form similar to the geometric deformations we use
 to test the stability of the background configuration, so the stiffness tensor' will determine at least in
 part the effective action of Goldstone modes to quadratic order. The fact that we observe flat directions implies that such modes are possible. The dependence of the energy on the transverse angular momentum  suggests that indeed the dispersion relation of the Goldstone modes is not simply linear in momentum.
 We hope to explore these questions and more in the future.

\section*{Acknowledgements}
We would like to thank Hirosi Ooguri and Chang-Soon Park for their
help in clarifying some aspects of \cite{Ooguri:2010kt,
Ooguri:2010xs}. The work of  C.H. and J.S. is partially supported by
the Israel Science Foundation (grant 1665/10). The work of J.S.  on
this project  was partially supported by the Einstein Center for
Theoretical Physics at the Weizmann Institute.

\appendix

\section{General deformation for a gauge-invariant energy density}\label{app:Ctilde}

 We can formally generalize the analysis of static deformations
to allow deformations that are time-dependent and involve the time
directions (such as boosts). One can use these methods to
derive an effective action for deformations. With this in mind, we will now consider a general deformation
of a theory with Abelian gauge fields.

The change of coordinates is
\begin{equation}
\tilde{x}^\mu=x^\mu-\xi^\mu(x).
\end{equation}
From
\begin{equation}
\partial_\nu\xi^\mu=\frac{\partial \tilde{x}^\alpha}{\partial x^\nu}\tilde{\partial}_\alpha \xi^\mu=(\delta^\alpha_\nu-\partial_\nu\xi^\alpha)\tilde{\partial}_\alpha \xi^\mu,
\end{equation}
we get
\begin{equation}
\partial_\nu\xi^\alpha=\tilde{\partial}_\nu\xi^\sigma(\delta_\sigma^\alpha -\tilde{\partial}_\sigma\xi^\alpha).
\end{equation}

The field strength is, to second order
\begin{equation}
\frac{\partial \tilde{x}^\alpha}{\partial x^\mu}\frac{\partial \tilde{x}^\beta}{\partial x^\nu}F_{\alpha\beta}=F_{\mu\nu}+\delta F_{\mu\nu}^{(1)}+\delta F_{\mu\nu}^{(2)},
\end{equation}
where
\begin{equation}
\delta F_{\mu\nu}^{(1)}=-\tilde\partial_\sigma\xi^\rho(\delta_\mu^\sigma F_{\rho\nu}+\delta_\nu^\sigma F_{\mu\rho}),
\end{equation}
and
\begin{equation}
\delta F_{\mu\nu}^{(2)}=\frac{1}{2} \tilde\partial_\sigma\xi^\rho\tilde\partial_\lambda\xi^\tau(\delta_\mu^\sigma \delta_\rho^\lambda F_{\tau\nu}+\delta_\nu^\sigma\delta_\rho^\lambda F_{\mu\tau}+(\sigma,\rho)\leftrightarrow (\lambda,\tau)).
\end{equation}
The Jacobian to second order is
\begin{align}
 &\left|\left|\frac{\partial x^\mu}{\partial \tilde{x}^\nu}\right|\right|= 1+J^{(1)}+J^{(2)},
\end{align}
where
\begin{equation}
J^{(1)}=\tilde\partial_\sigma\xi^\sigma,
\end{equation}
and
\begin{equation}
J^{(2)}=\frac{1}{2}\tilde\partial_\sigma\xi^\rho\tilde\partial_\lambda\xi^\tau(\delta_\rho^\sigma \delta_\tau^\lambda-\delta^\sigma_\tau\delta^\lambda_\rho).
\end{equation}
The energy to second order is then
\begin{align}
\notag &E_\Lambda=E+\int d^d x\, \left(J^{(1)}\cE+\frac{\delta \cE}{\delta F_{\mu\nu}} \delta F_{\mu\nu}^{(1)}\right)\\
&+\int d^d x\,\left(J^{(2)}\cE+J^{(1)}\frac{\delta \cE}{\delta F_{\mu\nu}} \delta F_{\mu\nu}^{(1)}+\frac{\delta \cE}{\delta F_{\mu\nu}} \delta F_{\mu\nu}^{(2)}+\frac{1}{2}\frac{\delta^2 \cE}{\delta F_{\mu\nu}\delta F_{\alpha\beta}} \delta F_{\mu\nu}^{(1)}\delta F_{\alpha\beta}^{(1)}\right).
\end{align}
We can now compute terms which are second order in the deformations,
\begin{align}
\delta E^{(2)} = \frac{1}{2}\int d^dx
\tC^{\alpha\gamma}_{\beta\delta}\d_\alpha\xi^\beta\d_\gamma\xi^\delta,
\end{align}
where $\tC^{ij}_{kl}=C^{ij}_{kl}$ is the  same tensor as in \eqref{Cgauge} (for $\delta\cE/\delta A_\mu=0$), but $\tC$ has also temporal indices:
\begin{align}
\tC^{\alpha\gamma}_{\beta\delta}=& \
\cE\left(\delta^\alpha_\beta\delta^\gamma_\delta-\delta^\alpha_\delta\delta^\gamma_\beta
\right)-\frac{\delta\cE}{\delta F_{\mu\nu}}\left(
F_{\mu\beta}\delta^\alpha_\nu\delta^\gamma_\delta+F_{\beta\nu}\delta^\alpha_\mu\delta^\gamma_\delta
+ (\alpha,\ \beta)\leftrightarrow (\gamma,\ \delta)
\right)\nonumber\\
&+\frac{\delta\cE}{\delta F_{\mu\nu}}\left(\delta_\mu^\alpha \delta_\beta^\gamma F_{\delta\nu}+\delta_\nu^\alpha\delta_\beta^\gamma F_{\mu\delta}+(\alpha,\beta)\leftrightarrow (\gamma,\delta)\right)
\nonumber\\
&+\frac{\delta^2\cE}{\delta F_{\mu\nu}\delta
F_{\rho\sigma}}\left(F_{\mu\beta}\delta^\alpha_\nu +F_{\beta\nu}\delta^\alpha_\mu\right) \left(F_{\rho\delta}\delta^\gamma_\sigma+F_{\delta\sigma}\delta^\gamma_\rho\right)~.
\end{align}

\section{Deformed configuration as force-free initial condition}\label{initialc}

The generic form of the background solution is
\begin{align}
&A_3= h_k \cos(k x_2+\varphi),\\
&A_4= h_k \sin(k x_2+\varphi),\\
& A_0= f_k.
\end{align}
We will modify this solution by a shift of the momentum and the phase,
\begin{equation}
k\to k+\delta k, \ \ \varphi\to\varphi+\delta\varphi.
\end{equation}
To leading order
\begin{align}
 &\delta A_3=a_3= \delta k\partial_k h_k \cos(k x_2+\varphi)-(x_2\delta k+\delta\varphi)\sin(k x_2+\varphi),\\
 &\delta A_4=a_4= \delta k\partial_k h_k \sin(k x_2+\varphi)+(x_2\delta k+\delta\varphi)\cos(k x_2+\varphi),\\
 &\delta A_0=a_0= \delta k\partial_k f_k.
\end{align}
We will restrict the change of the solution of wavelength $k$ to an interval $[-\pi/k,\pi/k]$. A change of the interval by $\delta k$ will modify the solution only by $\delta k^2$ terms, so we can neglect it.
\begin{align}\label{eq:dk}
\delta k= \delta k_0 \Delta(x_2), \ \ \delta\varphi=\delta\varphi(x_2,r),
\end{align}
where
\begin{equation}\label{eqdelta}
\Delta(x)\simeq \left(\Theta\left(x+\frac{\pi}{k}\right)-\Theta\left(x-\frac{\pi}{k}\right)\right),
\end{equation}
Here $\Theta(x)$ is a step function. We can regularize the solution by changing the step functions to
\begin{align}
&\left(\Theta\left(x+\frac{\pi}{k}\right)-\Theta\left(x-\frac{\pi}{k}\right)\right) \longrightarrow \frac{1}{2}\left( \tanh\left[M \left(x+\frac{\pi}{k}\right)\right]-\tanh\left[M\left(x-\frac{\pi}{k}\right)\right]\right),
\end{align}
with $M\gg 1$. In this case
\begin{equation}
\Delta(x)=\frac{1}{2}\left( \tanh\left[M \left(x+\frac{\pi}{k}\right)\right]-\tanh\left[M\left(x-\frac{\pi}{k}\right)\right]\right).
\end{equation}

Expanding the gauge fields in background plus fluctuations $F_{MN}+f_{MN}$, the linearized equations of motion are
\begin{equation}
\partial_M(\sqrt{-g} g^{MA} g^{NB} f_{AB})+\alpha \epsilon^{NABCD}F_{AB} f_{CD}=0.
\end{equation}
For the spatially modulated solution where $F_{r0}$ $F_{23}$, $F_{r3}$, $F_{24}$ and $F_{r4}$ are different from zero this leads to
\begin{align}
\label{eq1} 0&= \partial_M(\sqrt{-g} g^{MA} g^{22} f_{A2})+4\alpha h'(\cos(k x_2) f_{04}-\sin(k x_2)f_{03})-4\alpha A_0' f_{34},\\
\label{eq2} 0&=\partial_M(\sqrt{-g} g^{MA} g^{rr} f_{Ar})+4\alpha k h(\sin(k x_2)f_{04}+\cos(k x_2) f_{03}),\\
\label{eq3} 0&=\partial_M(\sqrt{-g} g^{MA} g^{33} f_{A3})+4\alpha A_0' f_{24}+4\alpha k h \cos(k x_2) f_{r0}+4\alpha h'\sin(k x_2)f_{02},\\
\label{eq4} 0&=\partial_M(\sqrt{-g} g^{MA} g^{44} f_{A4})-4\alpha A_0' f_{23}+4\alpha k h \sin(k x_2) f_{r0}-4\alpha h'\cos(k x_2) f_{02},\\
\notag 0&=\partial_M(\sqrt{-g} g^{MA} g^{00} f_{A0})-4\alpha k h(\sin(k x_2)f_{r4}+\cos(k x_2) f_{r3})\\ \label{eq5}
&-4\alpha h'(\cos(k x_2)f_{24}-\sin(k x_2) f_{23}).
\end{align}
For an ansatz where $a_2=0$ and there is no dependence on the $x^3$ and $x^4$ coordinates, the equations \eqref{eq1}, \eqref{eq2} and \eqref{eq5} are constraints depending only on first or zero time derivatives of the gauge potential. The other two equations, \eqref{eq3} and \eqref{eq4} are dynamical, they contain two time derivatives of $a_3$ and $a_4$. We want to study the time evolution of the system starting with an initial configuration similar to \eqref{eq:dk}. The constraint equations \eqref{eq1} and \eqref{eq2} are satisfied automatically. When we evaluate explicitly the third constraint \eqref{eq5}) there is a term proportional to $\delta k$
\begin{equation}
\delta k \left[(r^2 \partial_k f')'+4k h\partial_k h'+4kh'\partial_k h+4 h h' \right].
\end{equation}
This is the variation with respect to $k$ of the equation of motion for the background solution $f_k$, so it will vanish. The remaining contributions are
\begin{equation}
\frac{r}{H}\partial_k f \partial_2^2 \delta k+4 h'h(x_2 \partial_2\delta k+\partial_2 \delta\varphi)=0.
\end{equation}
This is satisfied if
\begin{equation}
\delta\varphi=\delta\varphi_0-x_2\delta k+\int dx_2 \delta k-\frac{r}{H}\frac{\partial_k f_k}{4h h'}\partial_2 \delta k.
\end{equation}
For \eqref{eqdelta},
\begin{align}
\notag \delta\varphi &=\delta\varphi_0-\delta k_0 x_2 \Delta(x_2)+\frac{\delta k_0}{2M}\log\left[ \frac{\cosh\left[ M\left(x_2+\frac{\pi}{k}\right)\right]}{\cosh\left[M\left(x_2-\frac{\pi}{k}\right)\right]}\right]\\
&-\frac{r}{H}\frac{\partial_k f_k}{8 h h'}M\delta k_0\left({\rm sech}^2\left[M\left(x_2+\frac{\pi}{k}\right)\right]-{\rm sech}^2\left[M\left(x_2-\frac{\pi}{k}\right)\right)\right].
\end{align}
The change in the phase of the solution is, up to terms that are exponentially localized around $x_2=\pm \pi/k$,
\begin{equation}
\delta\theta=\delta k x_2+\delta \varphi\simeq \delta\varphi_0+\frac{\delta k_0}{2M}\log\left[ \frac{\cosh\left[ M\left(x_2+\frac{\pi}{k}\right)\right]}{\cosh\left[M\left(x_2-\frac{\pi}{k}\right)\right]}\right].
\end{equation}
The asymptotic value is a constant
\begin{equation}
\delta\theta_{\pm\infty}=\lim_{x_2\to \pm\infty}\delta \theta=\delta\varphi_0 \pm \frac{\pi\delta k_0}{k}.
\end{equation}
One can see that $\delta\theta \sim \delta k_0 x_2+\delta\varphi_0$ in the interval $[-\pi/k,\pi/k]$ and
$\delta \theta\simeq \delta\theta_{\pm\infty}$ outside the interval. Therefore the initial configuration has the desired form, where the shift in $k$ is restricted to the interval, but the phase of the solution outside the interval is shifted in such a way that it is continuous in the limit $M\to \infty$. In the linearized approximation we can add three of these solutions in such a way that the asymptotic phase of the solution at positive and negative infinity is the same.  We simply add the solution that shifts $k$  by $\delta k_0$ in the interval $[-\pi/k,\pi/k]$ with the solutions that shift $k$ by $-\delta k_0/2$ in the intervals $[-3\pi/k,-\pi/k]$ and $[\pi/k,3\pi/k]$.

Up to exponentially localized terms in the boundaries of the interval, the initial configuration with $\delta k_0<0$ describes precisely the situation where a strip is stretched while the neighbouring strips are compressed. The force analisys tell us that in the regions where $\partial_k p <0$, the forces would tend to increase the deformation even more.



\begin{thebibliography}{99}


\bibitem{Deryagin:1992rw}
  D.~V.~Deryagin, D.~Y.~.Grigoriev and V.~A.~Rubakov,
  ``Standing wave ground state in high density, zero temperature QCD at large N(c),''
  Int.\ J.\ Mod.\ Phys.\ A {\bf 7}, 659 (1992).


\bibitem{Bringoltz:2009ym}
  B.~Bringoltz,
  ``Solving two-dimensional large-N QCD with a nonzero density of baryons and arbitrary quark mass,''
  Phys.\ Rev.\ D {\bf 79}, 125006 (2009)
  [arXiv:0901.4035 [hep-lat]].


\bibitem{Shuster:1999tn}
  E.~Shuster and D.~T.~Son,
  ``On finite density QCD at large N(c),''
  Nucl.\ Phys.\ B {\bf 573}, 434 (2000)
  [hep-ph/9905448].

\bibitem{Kharzeev:2010gd}
  D.~E.~Kharzeev and H.~-U.~Yee,
  ``Chiral Magnetic Wave,''
  Phys.\ Rev.\ D {\bf 83}, 085007 (2011)
  [arXiv:1012.6026 [hep-th]].

\bibitem{Basar:2010zd}
  G.~Basar, G.~V.~Dunne and D.~E.~Kharzeev,
  ``Chiral Magnetic Spiral,''
  Phys.\ Rev.\ Lett.\  {\bf 104}, 232301 (2010)
  [arXiv:1003.3464 [hep-ph]].

\bibitem{larkin:1964zz}
  A.~I.~larkin and Y.~N.~Ovchinnikov,
  ``Nonuniform state of superconductors,''
  Zh.\ Eksp.\ Teor.\ Fiz.\  {\bf 47}, 1136 (1964)
  [Sov.\ Phys.\ JETP {\bf 20}, 762 (1965)].

\bibitem{Fulde}
P.~Fulde R.~A.~Ferrell,\\
``Superconductivity in a Strong Spin-Exchange Field,''
  Phys.\ Rev.\ A\ {\bf 135}, 550 (1964)

\bibitem{Vojta}
 M.~Vojta,
``Lattice symmetry breaking in cuprate superconductors: stripes, nematics, and superconductivity,''
  Advances in Physics\ {bf 58}, 699 (2009)
[arXiv:0901.3145 [cond-mat.supr-con]]

\bibitem{Domokos:2007kt}
S.~K. Domokos and J.~A. Harvey, ``{Baryon number-induced Chern-Simons couplings
  of vector and axial-vector mesons in holographic QCD},''
  \href{http://dx.doi.org/10.1103/PhysRevLett.99.141602}{{\em Phys. Rev. Lett.}
  {\bfseries 99} (2007) 141602},
\href{http://arxiv.org/abs/0704.1604}{{\ttfamily arXiv:0704.1604 [hep-ph]}}.


\bibitem{Nakamura:2009tf}
S.~Nakamura, H.~Ooguri, and C.-S. Park, ``{Gravity Dual of Spatially Modulated
  Phase},'' \href{http://dx.doi.org/10.1103/PhysRevD.81.044018}{{\em Phys.
  Rev.} {\bfseries D81} (2010) 044018},
\href{http://arxiv.org/abs/0911.0679}{{\ttfamily arXiv:0911.0679 [hep-th]}}.



\bibitem{Bayona:2011ab}
C.~A.~B. Bayona, K.~Peeters, and M.~Zamaklar, ``{A non-homogeneous ground state
  of the low-temperature Sakai-Sugimoto model},''
  \href{http://dx.doi.org/10.1007/JHEP06(2011)092}{{\em JHEP} {\bfseries 06}
  (2011) 092},
\href{http://arxiv.org/abs/1104.2291}{{\ttfamily arXiv:1104.2291 [hep-th]}}.


\bibitem{Bergman:2011rf}
O.~Bergman, N.~Jokela, G.~Lifschytz, and M.~Lippert, ``{Striped instability of
  a holographic Fermi-like liquid},''
  \href{http://dx.doi.org/10.1007/JHEP10(2011)034}{{\em JHEP} {\bfseries 10}
  (2011) 034},
\href{http://arxiv.org/abs/1106.3883}{{\ttfamily arXiv:1106.3883 [hep-th]}}.


\bibitem{Iizuka:2013ag}
  N.~Iizuka and K.~Maeda,
  ``Stripe Instabilities of Geometries with Hyperscaling Violation,''
  arXiv:1301.5677 [hep-th].

\bibitem{Donos:2011ff}
A.~Donos and J.~P. Gauntlett, ``{Holographic helical superconductors},''
  \href{http://dx.doi.org/10.1007/JHEP12(2011)091}{{\em JHEP} {\bfseries 12}
  (2011) 091},
\href{http://arxiv.org/abs/1109.3866}{{\ttfamily arXiv:1109.3866 [hep-th]}}.

\bibitem{Donos:2012gg}
A.~Donos and J.~P. Gauntlett, ``{Helical superconducting black holes},''
\href{http://arxiv.org/abs/1203.0533}{{\ttfamily arXiv:1203.0533 [hep-th]}}.


\bibitem{Donos:2011qt}
A.~Donos, J.~P. Gauntlett, and C.~Pantelidou, ``{Spatially modulated
  instabilities of magnetic black branes},''
  \href{http://dx.doi.org/10.1007/JHEP01(2012)061}{{\em JHEP} {\bfseries 01}
  (2012) 061},
\href{http://arxiv.org/abs/1109.0471}{{\ttfamily arXiv:1109.0471 [hep-th]}}.


\bibitem{Donos:2011pn}
A.~Donos, J.~P. Gauntlett, and C.~Pantelidou, ``{Magnetic and electric AdS
  solutions in string- and M- theory},''
\href{http://arxiv.org/abs/1112.4195}{{\ttfamily arXiv:1112.4195 [hep-th]}}.


\bibitem{Donos:2012yu}
  A.~Donos, J.~P.~Gauntlett, J.~Sonner and B.~Withers,
  ``Competing orders in M-theory: superfluids, stripes and metamagnetism,''
  JHEP {\bf 1303}, 108 (2013)
  [arXiv:1212.0871 [hep-th]].


\bibitem{Donos:2013gda}
  A.~Donos and J.~P.~Gauntlett,
  ``Holographic charge density waves,''
  arXiv:1303.4398 [hep-th].


\bibitem{Takeuchi:2011uk}
S.~Takeuchi, ``{Modulated Instability in Five-Dimensional U(1) Charged AdS
  Black Hole with R**2-term},''
  \href{http://dx.doi.org/10.1007/JHEP01(2012)160}{{\em JHEP} {\bfseries 01}
  (2012) 160},
\href{http://arxiv.org/abs/1108.2064}{{\ttfamily arXiv:1108.2064 [hep-th]}}.


\bibitem{BallonBayona:2012wx}
  A.~Ballon-Bayona, K.~Peeters and M.~Zamaklar,
  ``A chiral magnetic spiral in the holographic Sakai-Sugimoto model,''
  JHEP {\bf 1211}, 164 (2012)
  [arXiv:1209.1953 [hep-th]].


\bibitem{Jokela:2012vn}
  N.~Jokela, G.~Lifschytz and M.~Lippert,
  ``Magnetic effects in a holographic Fermi-like liquid,''
  JHEP {\bf 1205}, 105 (2012)
  [arXiv:1204.3914 [hep-th]].


\bibitem{deBoer:2012ij}
  J.~de Boer, B.~D.~Chowdhury, M.~P.~Heller and J.~Jankowski,
  ``Towards a holographic realization of the Quarkyonic phase,''
  Phys.\ Rev.\ D {\bf 87}, 066009 (2013)
  [arXiv:1209.5915 [hep-th]].


\bibitem{Donos:2011bh}
A.~Donos and J.~P. Gauntlett, ``{Holographic striped phases},''
  \href{http://dx.doi.org/10.1007/JHEP08(2011)140}{{\em JHEP} {\bfseries 08}
  (2011) 140},
\href{http://arxiv.org/abs/1106.2004}{{\ttfamily arXiv:1106.2004 [hep-th]}}.


\bibitem{Ooguri:2010kt}
H.~Ooguri and C.-S. Park, ``{Holographic End-Point of Spatially Modulated Phase
  Transition},'' \href{http://dx.doi.org/10.1103/PhysRevD.82.126001}{{\em Phys.
  Rev.} {\bfseries D82} (2010) 126001},
\href{http://arxiv.org/abs/1007.3737}{{\ttfamily arXiv:1007.3737 [hep-th]}}.


\bibitem{Ooguri:2010xs}
H.~Ooguri and C.-S. Park, ``{Spatially Modulated Phase in Holographic
  Quark-Gluon Plasma},''
  \href{http://dx.doi.org/10.1103/PhysRevLett.106.061601}{{\em Phys. Rev.
  Lett.} {\bfseries 106} (2011) 061601},
\href{http://arxiv.org/abs/1011.4144}{{\ttfamily arXiv:1011.4144 [hep-th]}}.



\bibitem{Donos:2012wi}
  A.~Donos and J.~P.~Gauntlett,
  ``Black holes dual to helical current phases,''
  Phys.\ Rev.\ D {\bf 86}, 064010 (2012)
  [arXiv:1204.1734 [hep-th]].


\bibitem{Rozali:2013ama}
  M.~Rozali, D.~Smyth, E.~Sorkin and J.~B.~Stang,
  ``Striped Order in AdS/CFT,''
  arXiv:1304.3130 [hep-th].

\bibitem{Withers:2013kva}
  B.~Withers,
  ``The moduli space of striped black branes,''
  arXiv:1304.2011 [hep-th].

\bibitem{Donos:2013wia}
  A.~Donos,
  ``Striped phases from holography,''
  JHEP {\bf 1305}, 059 (2013)
  [arXiv:1303.7211 [hep-th]].


\bibitem{Iizuka:2012pn}
  N.~Iizuka, S.~Kachru, N.~Kundu, P.~Narayan, N.~Sircar, S.~P.~Trivedi and H.~Wang,
  ``Extremal Horizons with Reduced Symmetry: Hyperscaling Violation, Stripes, and a Classification for the Homogeneous Case,''
  JHEP {\bf 1303}, 126 (2013)
  [arXiv:1212.1948 [hep-th]].

\bibitem{Rozali:2012es}
  M.~Rozali, D.~Smyth, E.~Sorkin and J.~B.~Stang,
  ``Holographic Stripes,''
  arXiv:1211.5600 [hep-th].


\bibitem{Albash:2008eh} 
  T.~Albash and C.~V.~Johnson,
  ``A Holographic Superconductor in an External Magnetic Field,''
  JHEP {\bf 0809}, 121 (2008)
  [arXiv:0804.3466 [hep-th]].


\bibitem{Albash:2009ix} 
  T.~Albash and C.~V.~Johnson,
  ``Phases of Holographic Superconductors in an External Magnetic Field,''
  arXiv:0906.0519 [hep-th].

\bibitem{Albash:2009iq} 
  T.~Albash and C.~V.~Johnson,
  ``Vortex and Droplet Engineering in Holographic Superconductors,''
  Phys.\ Rev.\ D {\bf 80}, 126009 (2009)
  [arXiv:0906.1795 [hep-th]].

\bibitem{Montull:2009fe} 
  M.~Montull, A.~Pomarol and P.~J.~Silva,
  ``The Holographic Superconductor Vortex,''
  Phys.\ Rev.\ Lett.\  {\bf 103}, 091601 (2009)
  [arXiv:0906.2396 [hep-th]].


\bibitem{Keranen:2009vi} 
  V.~Keranen, E.~Keski-Vakkuri, S.~Nowling and K.~P.~Yogendran,
  ``Dark Solitons in Holographic Superfluids,''
  Phys.\ Rev.\ D {\bf 80}, 121901 (2009)
  [arXiv:0906.5217 [hep-th]].


\bibitem{Maeda:2009vf} 
  K.~Maeda, M.~Natsuume and T.~Okamura,
  ``Vortex lattice for a holographic superconductor,''
  Phys.\ Rev.\ D {\bf 81}, 026002 (2010)
  [arXiv:0910.4475 [hep-th]].


\bibitem{Bu:2012mq}
  Y.~-Y.~Bu, J.~Erdmenger, J.~P.~Shock and M.~Strydom,
  ``Magnetic field induced lattice ground states from holography,''
  JHEP {\bf 1303}, 165 (2013)
  [arXiv:1210.6669 [hep-th]].

\bibitem{Kaplunovsky:2012gb}
  V.~Kaplunovsky, D.~Melnikov and J.~Sonnenschein,
  ``Baryonic Popcorn,''
  JHEP {\bf 1211}, 047 (2012)
  [arXiv:1201.1331 [hep-th]].

\bibitem{Kaplunovsky:2013iza}
  V.~Kaplunovsky and J.~Sonnenschein,
  ``Dimension Changing Phase Transitions in Instanton Crystals,''
  arXiv:1304.7540 [hep-th].
\bibitem{Vegh:2013sk}
  D.~Vegh,
  ``Holography without translational symmetry,''
  arXiv:1301.0537 [hep-th].
\bibitem{Hartnoll:2012rj}
  S.~A.~Hartnoll and D.~M.~Hofman,
  ``Locally Critical Resistivities from Umklapp Scattering,''
  Phys.\ Rev.\ Lett.\  {\bf 108}, 241601 (2012)
  [arXiv:1201.3917 [hep-th]].

\bibitem{Horowitz:2012ky}
  G.~T.~Horowitz, J.~E.~Santos and D.~Tong,
  ``Optical Conductivity with Holographic Lattices,''
  JHEP {\bf 1207}, 168 (2012)
  [arXiv:1204.0519 [hep-th]].

\bibitem{Horowitz:2012gs}
  G.~T.~Horowitz, J.~E.~Santos and D.~Tong,
  ``Further Evidence for Lattice-Induced Scaling,''
  JHEP {\bf 1211}, 102 (2012)
  [arXiv:1209.1098 [hep-th]].

\bibitem{Donos:2012js}
  A.~Donos and S.~A.~Hartnoll,
  ``Metal-insulator transition in holography,''
  arXiv:1212.2998 [hep-th].

\bibitem{Liu:2012tr}
  Y.~Liu, K.~Schalm, Y.~-W.~Sun and J.~Zaanen,
  ``Lattice Potentials and Fermions in Holographic non Fermi-Liquids: Hybridizing Local Quantum Criticality,''
  JHEP {\bf 1210}, 036 (2012)
  [arXiv:1205.5227 [hep-th]].

\bibitem{Gibbons:1997xz}
  G.~W.~Gibbons,
  ``Born-Infeld particles and Dirichlet p-branes,''
  Nucl.\ Phys.\ B {\bf 514}, 603 (1998)
  [hep-th/9709027].

\bibitem{Domokos:2013xqa}
  S.~K.~Domokos, C.~Hoyos and J.~Sonnenschein,
  ``Deformation Constraints on Solitons and D-branes,''
  arXiv:1306.0789 [hep-th].


\bibitem{Donos:2013cka}
  A.~Donos and J.~P.~Gauntlett,
  ``On the thermodynamics of periodic AdS black branes,''
  arXiv:1306.4937 [hep-th].

\bibitem{Landau}
 L.~D.~Landau and E.~M.~Lifshitz,
 `` Theory of Elasticity,''
 Course of Theoretical Physics, Volume 7 Pergamon Press, 1975


\bibitem{Sakai:2004cn}
  T.~Sakai and S.~Sugimoto,
  ``Low energy hadron physics in holographic QCD,''
  Prog.\ Theor.\ Phys.\  {\bf 113}, 843 (2005)
  [hep-th/0412141].

\bibitem{Aharony:2006da}
  O.~Aharony, J.~Sonnenschein and S.~Yankielowicz,
  ``A Holographic model of deconfinement and chiral symmetry restoration,''
  Annals Phys.\  {\bf 322}, 1420 (2007)
  [hep-th/0604161].


\bibitem{Karch:2008uy}
  A.~Karch, A.~O'Bannon and E.~Thompson,
  ``The Stress-Energy Tensor of Flavor Fields from AdS/CFT,''
  JHEP {\bf 0904}, 021 (2009)
  [arXiv:0812.3629 [hep-th]].


\bibitem{Papadimitriou:2005ii} 
  I.~Papadimitriou and K.~Skenderis,
  ``Thermodynamics of asymptotically locally AdS spacetimes,''
  JHEP {\bf 0508}, 004 (2005)
  [hep-th/0505190].


\bibitem{Gubser:2008wv}
  S.~S.~Gubser and S.~S.~Pufu,
  ``The Gravity dual of a p-wave superconductor,''
  JHEP {\bf 0811}, 033 (2008)
  [arXiv:0805.2960 [hep-th]].

\bibitem{Ammon:2008fc}
  M.~Ammon, J.~Erdmenger, M.~Kaminski and P.~Kerner,
  ``Superconductivity from gauge/gravity duality with flavor,''
  Phys.\ Lett.\ B {\bf 680}, 516 (2009)
  [arXiv:0810.2316 [hep-th]].

\bibitem{Nielsen:1975hm}
  H.~B.~Nielsen and S.~Chadha,
  ``On How to Count Goldstone Bosons,''
  Nucl.\ Phys.\ B {\bf 105}, 445 (1976).

\bibitem{Watanabe:2011ec}
  H.~Watanabe and T.~Brauner,
  ``On the number of Nambu-Goldstone bosons and its relation to charge densities,''
  Phys.\ Rev.\ D {\bf 84}, 125013 (2011)
  [arXiv:1109.6327 [hep-ph]].

\bibitem{Amado:2013xya}
  I.~Amado, D.~Arean, A.~Jimenez-Alba, K.~Landsteiner, L.~Melgar and I.~S.~Landea,
  ``Holographic Type II Goldstone bosons,''
  arXiv:1302.5641 [hep-th].

\end{thebibliography}
\end{document}